\documentstyle[12pt]{article}

\def\singlespace{\baselineskip 12pt}
\def\doublespace{\baselineskip 16pt}
\def\be{\begin{equation}}
\def\ee{\end{equation}}
\def\de{\hbox{d}}

\newcommand{\spose}[1]{\hbox to 0pt{#1\hss}}
\newcommand{\lta}[0]{{\mathrel{\spose{\lower 3pt\hbox{$\mathchar"218$}}
     \raise 2.0pt\hbox{$\mathchar"13C$}}}}
\newcommand{\gta}[0]{{\mathrel{\spose{\lower 3pt\hbox{$\mathchar"218$}}
     \raise 2.0pt\hbox{$\mathchar"13E$}}}}

\newcommand{\msun}[0]{{M_\odot}}
\newcommand{\au}[0]{{\,\hbox{AU}}}
\newcommand{\kpc}[0]{{\,\hbox{kpc}}}

\newcommand{\kms}[0]{{\,\hbox{km/s}}}
\newcommand{\scrR}[0]{{\cal R}}

\begin{document}
\singlespace
\title{Truncation of the Binary\\
Distribution Function in Globular \\
Cluster Formation}
\author{E. Vesperini$^{1,2}$ and David F. Chernoff$^2$\\
$^1$ Scuola Normale Superiore,\\
Piazza dei Cavalieri 7, 56126-I, Pisa\\
$^2$ Department of Astronomy, Space Science Building, Cornell University,\\
14853 Ithaca (New York) USA}
\date{}
\maketitle
\clearpage
\begin{abstract}

We investigate a population of primordial binaries during the initial
stage of evolution of a star cluster.  For our calculations we assume
that equal mass stars form rapidly in a tidally truncated gas cloud,
that $\sim10$\% of the stars are in binaries and that the resulting
star cluster undergoes an epoch of violent relaxation.  We study the
collisional interaction of the binaries and single stars, in
particular, the ionization of the binaries and the energy exchange
between binaries and single stars.  We find that for large $N$ systems
($N > 10^3$) even the most violent beginning leaves the binary
distribution function largely intact. Hence, the binding energy
originally tied up in the cloud's protostellar pairs is preserved
during the relaxation process and the binaries
are available to interact at later
times within the virialized cluster.

\end{abstract}
Subject headings: clusters:globular, stars: binaries, stars: stellar dynamics\\
\clearpage
\doublespace
\section{Introduction}

Recent observational searches suggest that the frequency of primordial
binaries in globular clusters may reach $\approx 10\%$ (see Hut at al.
1992 for a review).  The inferred abundance of binaries is
sufficiently large that both dynamical and collisional consequences
are important.  Several different treatments conclude that binaries
are effective in halting core collapse, supporting the core and/or
driving an expansion phase (Goodman \& Hut 1989, McMillan et al.
1990, 1991, Gao et al.  1991, Heggie \& Aarseth 1992).  The abundance
and binding energy distribution of the binaries have a direct impact
on observable characteristics of globular clusters such as the size of
the core radius (Vesperini \& Chernoff 1994). The inferred abundance
of binaries is consistent with that needed by stellar reaction
pathways to account for the population of millisecond pulsars in the
cluster system (Sigurdsson and Phinney 1994).

The fraction of binaries observed among field stars ($>70\%$ see,
e.g., Abt 1983, Duquennoy \& Mayor 1991, and Reipurth \& Zinnecker
1993 for the frequency of binary stars among pre-main sequence stars)
is larger than the fraction, hitherto detected, in clusters. The size
of the difference may be partially due to observational biases:
binaries should segregate to the center of the cluster and may be
difficult to find on account of stellar crowding.  Or
fewer binaries may have formed originally, some may be depleted by
collisional effects and some visible members may be swapped by
exchange reactions for heavy, degenerate and invisible remnants. Many
lines of inquiry (reviewed by Hut et al. 1992) are being pursued to
characterize the present binary content and binary binding energy
distribution within clusters.

The key question we will address is whether it is possible for
binaries, assumed to be primordial, to survive the birth of the
cluster.  The route for the formation of a bound cluster of stars is
not well-understood but it is plausible that a hydrostatically
supported gas cloud is a direct precursor to the stellar cluster. Such
a cloud might form by cooling instability in the Galactic halo (Fall
\& Rees 1985) or in the cooling region behind a shock front after two
gas clouds collide (MacLow \& Shull 1986, Shapiro \& Kang 1987). The
narrowness of the main-sequence in Galactic globular clusters is
evidence that the epoch of star formation is brief. If star formation
proceeds rapidly compared to the gravitational free-fall time, the
resulting collection of stellar objects forms far out of virial
equilibrium and a period of violent relaxation ensues. Stars exchange
energy through fast, large-scale variation of the gravitational
potential.  Although the collapse is dissipationless, the gross
properties (radius, velocity dispersion) of the system undergo damped
oscillations until a virialized state is reached. During the intervals
of high stellar density and high velocity dispersion collisional
interactions occur with increased rate.

In this paper we will investigate the fate of a population of
primordial binaries, paying particular attention to the process of
disruption of binaries.  Since the total binary binding energy can
easily exceed the smooth, large-scale gravitational potential energy
of the stars of the cluster, binary heating and softening are also
potentially important processes to address.  Although we treat only
point-like stellar objects and assume all gas is removed at the
conclusion of star formation, our conclusions have some applicability
to the frequency of collisional interactions of stars with gaseous
disks.  Elsewhere, we will consider the implications for a scenario in
which stellar disks are common once violent relaxation begins.

To address the issues, we explore the rate of collisional interactions
between single stars and binaries.  We determine how the binary
distribution function is altered including the fraction of binaries
destroyed and identify a characteristic binding energy below which the
initial distribution is truncated.  To anticipate our results to some
extent, we find that collisional effects during the epoch of violent
relaxation are small in the following sense: in the final, virialized
state only {\it soft} binaries have been strongly effected (i.e.
heated or ionized) when the number of stars $N$ is as large as it
typically is for a globular cluster. We explore the scaling of this
conclusion with $N$.

The scheme of the paper is the following:
in \S 2 we estimate the binary destruction that occurs during the
violent relaxation; we discuss the rationale for carrying out an
N-body experiment;
in \S 3 we describe several tests of the N-body code designed to
verify that it performs adequately in all regimes of interest;
in \S 4 we describe the N-body runs and analyze the results;
in \S 5 we report a refined analytic estimate of the change in the
binary distribution function. We are able to explain most of the
significant trends in the N-body simulations and to verify the key role
that ionization plays in the numerical experiment.
In \S 6 we summarize the main conclusions and describe future work.

\section{Violent relaxation and binary destruction: a
simplified model}

Let $\xi \equiv T/\vert V\vert$ be the initial ratio of kinetic to
potential energy of the system.  For the estimates in this section, we
represent a stellar cluster by a uniform, spherically symmetric
density distribution of radius $R$.  If $\xi < 0.5$ the system
collapses and oscillates with decreasing amplitude as it
virializes. The first cycle of collapse and re-expansion (``bounce'')
produces the densest central conditions. The bounce occurs when the
kinetic energy associated with the particles'
dispersion becomes comparable to the potential
energy. $N$-body calculations show
that the discreteness may play a role in determining the cluster's size
at maximum collapse $R_{mc}$ (Aarseth et al.  1988).  If the initial
velocity dispersion exceeds a given amount ($\xi \gta N^{-1/3}$) then
the minimum size scales like $R_{mc} \sim \xi R_i$.
If the system is
very cold, so that $\xi \lta N^{-1/3}$, and if the
initial positions are randomly realized for a uniform density
distribution then the discrete fluctuations
induce a background of perturbations and $R_{mc} \sim N^{-1/3} R_i$.

If no star gains enough energy to escape and if the
cluster remains homogeneous, then the virialized state satisfies
energy conservation so that
\be
-{3 \over 5} {GM^2 \over R_i}(1-\xi)=-{3 \over 10} {GM^2 \over R_f}
\label{eq:ec}
\ee
where $M$ and $R_i$ are respectively the mass and the initial radius
of the system. Hence, the final radius is
\be
R_f= {R_i \over 2(1-\xi)} .
\label{eq:rfinal}
\ee
If some of the stars escape from the system, changing the mass
by $\Delta M = M_f - M_i$ and the energy by $\Delta E = E_f - E_i$
(where final values are indicated by ``f'') but leaving the cluster
homogeneous, then energy conservation implies
\be
-{3 \over 10} {GM_f^2 \over R_f}=-{ 3 \over 5} {GM_i^2 \over R_i}(1-\xi)
+\Delta E .
\ee
The final radius is
\be
R_f=  {R_i \over 2(1-\xi)}{(1+\Delta M/M_i)^2 \over (1+\Delta E/E_i)} .
\label{eq:fs}
\ee
If the system cools ($\Delta E <0$) and/or loses mass ($\Delta M < 0$)
then the final size is smaller than the case in which there is no
change of energy; if the system heats ($\Delta E >0$) but does not
lose mass, the opposite is true.

For a system of single stars, van Albada (1982) showed that the
magnitude of the mass and energy changes during violent relaxation
is correlated to how cold the system is at the beginning. Small $\xi$ implies
violent, large amplitude oscillations in the cluster and large
$|\Delta M|$ and $|\Delta E|$. When binaries are present and suffer
collisional interactions within the cluster, the gain or loss of their
binding energy is accompanied by changes in stellar translational energy.

The amount of energy tied up in the binary population can be
substantial. Let the number of single stars, binary systems and total
stars be $N_s$, $N_b$ and $N = N_s + 2N_b$, respectively.  The ratio
of the binary internal binding energy to large-scale cluster
binding energy ($3GN^2m^2/5R$) is
$\alpha = (5f_b/6N) (R \langle 1/a \rangle)$, where
$f_b = N_b/N$ is the fraction of binaries, $m$ is the mass per star
and $\langle 1/a \rangle$ is the cluster average of the inverse binary
semi-major axis.
For a tidally limited cluster, on a circular orbit of
Galactocentric radius $R_g$ with rotation velocity $v_{rot}$,
\be
\alpha = 114 f_b
\langle {\au \over a} \rangle
\left( N \over 10^5 \right)^{-2/3}
\left( {R_g \over 10 \kpc} { 220 \kms \over v_{rot} } \right)^{2/3}
\left( m \over \msun \right)^{1/3} .
\ee
For a newly formed cluster, the most uncertain element in the above
estimate is probably $\langle 1/a \rangle$.
The field binary distribution is peaked
about a period $P \sim 10^{4.8}$ d; the period distribution $dN_b(P)/d
\log P$ varies like a Gaussian with $\sigma_{\log P}=2.3$ (Duquennoy
\& Mayor 1991). The peak of the field star distribution corresponds to
$a \sim 30 \au$ but the quantity $\langle 1/a \rangle$
is effectively dominated by
binaries of the smallest separation.  For a cluster born with
field-like binaries, taking $a$ to be of order $30 \au$ and $f_b \sim
0.1$ (as observed today in globulars, rather than $f_b \sim 0.6$
as observed in the
field, Duquennoy \& Mayor 1991 and Reipurth \& Zinnecker
1993) gives an ultra-conservative lower limit for $\alpha$ of $0.4$.
A more plausible estimate is that $a$ extends to $\sim 1 \au$ (as
observed in the field population, and as required by millisecond
pulsar production pathways that involve binary mass transfer),
implying $\alpha \sim 11$.

For the available binding energy to play a dynamical role in the
cluster evolution, binaries must interact collisionally.  Collisional
interactions occur most rapidly at the point of maximum
contraction of the cluster when the background stellar density and
velocity dispersion are largest.  In the estimates here we
will focus on ionization, however, our $N$-body treatment is
completely general (\S 3 and 4) and includes all the point-like
interactions of single stars and binaries. We characterize the binary
destruction by the ``cutoff'' semi-major axis $a_{cut}$ which
corresponds to the smallest value of the initial semi-major axis $a$
for which the probability of destruction is of order unity. Ionization
in single encounters and cumulative energy changes by successive
encounters both contribute to changes in the form of binary binding
energy distribution and in the cluster's translational energy.

The change in the number density of binaries ($n_b$) by ionizing collisions
with single stars of density $n_s$ is
\be
{\de n_b \over \de t}= - n_s n_b\langle \sigma v\rangle
\label{eq:ionrate}
\ee
where $\langle \sigma v \rangle$ is the rate coefficient, averaged
over the relative velocity distribution. If the single
stars and binaries both have Maxwellian velocity distributions,
the rate coefficient may be written
\be
\langle \sigma v \rangle = \pi a^2 V_{th} \scrR (x)
\ee
where $\scrR (x)$ is a function of the hardness factor $x \equiv
\epsilon/mv_s^2$ ($\epsilon$ is the binary binding energy $Gm^2/2a$
and $v_s$ is the single star, one-dimensional velocity dispersion)
given by fits to numerical simulations (Hut \& Bahcall 1983) as
\be
\scrR (x)={ 1.64  \over
       \left(1 + 0.2A/x \right)
       \left(1 + \hbox{exp}\left[{x/A}\right]\right) } ,
\ee
$V_{th} = 3(A/2)^{1/2}v_s$ is the dispersion in relative velocity and $A$
is a constant. For the case of energy equipartition $A=1$; for
velocity equipartition $A=4/3$. The process of violent relaxation
leaves the stars in the cluster close to velocity equipartition.

Let the half-mass radius at the point of maximum contraction
be $R_{h,mc}$.  The density $\rho \sim M R_{h,mc}^{-3}$, the
velocity $v_s \sim (G M/R_{h,mc})^{1/2}$ and the binary hardness $x
\sim R_{h,mc}/aN$.  The lapse of time spent by the system in its
maximally contracted form is proportional to the dynamical time
$t_{dyn,mc} \sim (G\rho_{mc})^{-1/2}$. The fractional decrease in
binaries of size $a$ is
\begin{eqnarray}
\delta (a)  & \equiv & {1 \over n_b}{\de n_b \over \de t} t_{dyn,mc}  \\
       & = &      C \left( a^2 N \over R_{h,mc}^{2} \right)
  \scrR \left(B \left[ R_{h,mc} \over a N \right] \right)
\end{eqnarray}
where $C \simeq 1.16$ and $B \simeq 1.56$ are numerical coefficients
obtained by keeping all the appropriate factors in the above
estimates. If the binary is soft at the point of maximum collapse, then
we have the following simple expression for the fractional
decrease
\be
\delta = 5.56 {a \over R_{h,mc}} .
\label{eq:delta}
\ee
A more complete analysis is possible and is given in Appendix A.
 The cutoff semi-major axis (for
which $\delta \sim 1$) is closely tied to the minimum scale of the
collapse.

Aarseth et al. (1988) characterized the minimum scale
for collapse of a cold spherical system without angular momentum. In
basic agreement with their results our numerical work (in succeeding
sections) shows that the system collapses to a minimum scale set by
root-N fluctuations in the initial conditions. The theory predicts
$R_{h,mc} \propto N^{-1/3}$ and we find
\be
R_{h,mc} = {R_{h,i} \over 1.13 N^{0.36}} .
\ee
For the homogeneous system of interest here, $R_i=2^{1/3}R_{h,i}$, and
$$
R_{h,mc} = {R_i \over 1.42 N^{0.36}}
$$
which implies
\be
a_{cut} \sim 0.13 {R_i \over N^{0.36}} .
\label{eq:acutoff}
\ee
The hardness of this binary, measured in the final virialized system,
assuming no mass or energy loss during the epoch of violent
relaxation, and making the approximation $\xi = 0$ is
\be
x_{cut} \sim {9.6 \over N^{0.64}} .
\ee
{\it If $N \gta 10^2$ only soft binaries are
ionized}. The simple analysis indicates that violent
relaxation truncates the binary distribution at a characteristic soft
scale.

The estimates based on the simple model are potentially misleading
because the model has the following shortcomings:

\noindent (1) The model
assumes homogeneity while numerical simulations (see, e.g., van
Albada 1982, Aarseth et al. 1988) show that the collapse of an
initially homogeneous cluster gives rise to a core-halo structure when
the point of maximum contraction is reached and that strong density
variations persist in the final state. Since the collisional reaction
rates depend sensitively on density it is not clear how the
estimates will be altered.

\noindent (2) The rate coefficient for the model estimate of ionization is
based on Maxwellian velocity distributions.  Since there exist
strongly ordered motions during the initial phase of violent
relaxation, not thermal distributions, the rate of binary destruction
may be inaccurate.

\noindent (3) In a cold initial state, particles are correlated over long
distances; bound binaries, triples, etc. are present in the initial
conditions. The rate of destruction of these structures cannot be
described with reaction rate coefficients for free particles impinging
on a binary as assumed in the model.

$N$-body simulations are described in \S 3 and \S 4 which are free of
these shortcomings. Qualitatively, they validate the conclusion of the
simple model that only soft binaries are ionized. Quantitatively, they
give a more accurate description of the processes that are important
in the evolution of the binary distribution function.

\section{Tests of the numerical code used}

Simulations have been carried out by a special $N$-body code designed
for collisional problems (Jayaraman \& Chernoff 1992).  Forces are
calculated using an oct tree and a low-order multipole expansion of
the particles within a cell (following Salmon 1991). Particles have
independent time steps; polynomial approximations are used to specify
positions and velocities within given time-intervals. Updated
polynomials are found using a predictor-corrector integration scheme
with accuracy O($\Delta t^{3.5}$). The tree's structure is exact with
respect to current particle positions (i.e. it is not allowed to
deform as in the treatment of McMillan and Aarseth 1993).  The
accuracy of the integration is controlled by two tolerance parameters,
one for the integration step size (for both relative and absolute
error controls) and one for the neighbor sphere size.  Well-isolated,
bound stellar pairs are treated in the following special manner.
The internal degrees of
freedom are regarded as unperturbed Kepler orbits and the pairs
interact with the rest of the system as single entities having the
total mass and momentum of the constitutents.  All non-isolated stellar
pairs are treated as individual stars. The classification into
isolated and non-isolated pairs is updated on each timestep.
Close-encounters between singles and/or binaries are identified making
use of the tree structure. When a close-encounter is detected the
particles involved are advanced on a collision-based timescale, which
is typically much shorter than the timestep a single star would take
in a collisionless simulation. Since each object is advanced
independently, the extra cost for handling close encounters in this
way is not prohibitive.  The code maintains collisionless equilibria
including Plummer laws, King models and polytropes for many
characteristic dynamical times.

Several simulations were designed to test critical aspects necessary for
the simulations of interest. In the first set of tests the collapse
of a homogeneous sphere of particles was followed. The evolution provides
a good illustration of the behavior of the system in the absence of
primordial binaries and allows calibration of errors and comparison with
previous results.

Table 1 summarizes the results of three runs with $N=1000$.  The stars
are distributed homogeneously within a sphere with virial ratios $\xi
= 0.25$, $0.1$ and $0.01$.  The particle velocities are drawn from a
flat 3d distribution function whose maximum value is determined by
$\xi$.

The units of all runs have been chosen so that $G=1$, the total
initial mass of the system $M_i=1$ and the initial radius of the
system $R_i=1$. The timescale for the homogeneous sphere to
collapse is
\be
t_{ff}=\sqrt{{3\pi \over 32 G\rho_{i}}} = 1.11
\label{eq:ff}
\ee
The system is already close to the state of virial equilibrium at
$t=1.3$; runs were stopped at $t=2$.  Figure 1 provides snapshots of
the mass density profiles at $t=0.5$, $1.1$, $1.5$, and $2$ for the
case $\xi = 0.01$. In agreement with van Albada (1982) the snapshots
clearly show the development of a core-halo structure during the
collapse. The central density reaches its maximum value at $t \simeq
t_{ff}$ as expected.

Table 1 shows, in agreement with van Albada (1982; see his table [1]),
the physical trends that clusters with smaller $\xi$ have (i) a larger
fraction of escaping stars, (ii) a greater energy change in the bound
stars and (iii) a larger ratio of final to initial central densities,
$\rho_{c,f}/\rho_{c,i}$. In agreement with the analytic model of \S 2
the final to initial half-mass radius, $R_{h,f}/R_{h,i}$ may be
predicted from the observed mass and energy changes.  Table 1 also
gives an indication of the size of the numerical errors.  For
all runs herein, the numerical parameters have been adjusted to give relative
energy errors $\lta 1$\% over the entire course of the calculation.

The second test assesses the ability of the code to track collisional
encounters not only during the phase of violent relaxation but also
over relaxation-length timescales.  The single
and binary stars are homogeneously distributed in a sphere with $\xi =
0.01$ as above. Here $N=1000$ with $N_s = 500$ and $N_b = 250$. The
binaries have semi-major axis $a \simeq 9 \times 10^{-4}$ and zero
eccentricity. In the virialized state, assuming no energy or mass
loss, the binary's hardness is $x = 5/4Na \sim 1.4$.

The evolution of the system was followed well beyond the end of
violent oscillations until $t=8$ (approximately $\sim 7 t_{ff}$).
This corresponds to $\sim 100 t_{dyn}$, where we employ the
definition of Spitzer and Hart (1971)
\be
t_{dyn}= { {1.58 R_h^{3/2}} \over M^{1/2} } ,
\ee
evaluated in the virialized state. (Specifically, at $t = 2$, the
measured value of the $R_h$ implies $t_{dyn} \approx 0.07 t_{ff}$.)
The half-mass relaxation time $t_{rh} = N t_{dyn}/26 \log(0.4N) \sim
15 t_{dyn}$. The long period of quiescent evolution permits a
significant number of binary-single interactions.

Figure 2 shows the energy budget of the stars within the systems split
into four separate categories (analogous plots are shown in Heggie \&
Aarseth 1992) (i) into internal energy (i.e. negative binding energy
of a binary) versus external energy (i.e. translational energy of a
single star or of a binary center of mass plus large-scale potential
energy) contributions and (ii) into retained (bound to the cluster)
versus escaping particles. The separate quantities plotted are labeled
$E_{int}$, $E_{int,esc}$, $E_{ext}$, and $E_{ext,esc}$. The total
energy of the cluster is $E_{int} + E_{ext}$; the conserved energy is
the sum of all four components.

Evolution proceeds through several distinct phases.  {\it Infall}
occurs for $t < t_{ff}$; there are no escaping particles yet.  The
binding energy associated with the binaries increases slightly (line
of medium dashs) because the binaries are {\it hard} relative to the
cold, background stellar distribution. The interactions that take
place during this phase depend, in part, upon pre-existing bound
configurations that have the opportunity to collapse and interact
slightly before $t_{ff}$. The collisional interactions lead to a small
degree of binary hardening and a concomitant heating of the stellar
distribution, seen as a slight increase in the external energy (line
of short dashes).

Mass {\it ejection} occurs near $t_{ff}$ at the point of maximum
contraction.  This is the largest change that occurs over the course
of the entire simulation. For the binary distribution realized, the
external energy dominates the internal energy for the escaping matter,
but this need not be true in general.  The figure clearly indicates
the corresponding decrease in the external energy of the matter that
remains bound.

After the ejection has occurred, the binaries within the cluster {\it
slowly evolve}. They soften (their internal binding energy decreases)
while the external energy of the cluster decreases. There is little
discernible mass or energy loss during this phase. (Note, no tidal
field was imposed on the cluster.) Due to the ejection of mass and
energy during the violent relaxation, the hardness of the binaries is
$\sim 0.4$, somewhat less than the value $1.4$ anticipated on the
basis of the homogeneous model. Since $x \lta 1$ the binaries soften.

The main check provided by this test is energy conservation: over the
course of the entire run the relative energy conservation is
$\sim 10^{-3}$.

\section{$N$-body simulations}

\subsection{Gross Changes}

In this section we describe a set of N-body simulations designed to
answer the related questions: (1) Is cluster evolution grossly altered
by the presence of the primordial binaries during the period of
violent relaxation? (2) How is the binary distribution modified by
collisional interactions? (3) Does significant mass segregation of the
binary population occur? We take the initial stellar system to be
spherical, homogeneous and cold ($\xi=0.01$). We make these choices in
an attempt to maximize the central density and the destruction and/or
collisional interactions of the binaries. In another paper we will
discuss the role of inhomogeneities induced by a
tidal field in limiting the degree of compression.

The adopted units are $G=1$, total initial mass is $M_i=1$, initial
radius $R_i=1$. We first generate a set of initial conditions with
single stars (mass $m$) and binaries composed of stars (of mass $m$)
and refer to this as the ``normal'' case.  For these runs, the initial
binary fraction is $f_b = 0.05$ and all binaries begin with circular
orbits.  The distribution function for the binding energy $\epsilon$
is logarithmic, i.e.  $dN = f(\epsilon) d\epsilon \propto
d\epsilon/\epsilon$.  The selected range of binary energies
corresponds approximately to $x \in [0.13,1.3]$ in a homogeneous,
virialized state, assuming no energy or mass loss and encompasses the
critical cutoff semi-major axis $a_{cut}$. All calculations are
carried out to $t=1.3$ at which point the system is close to virial
equilibrium. We also consider two variants of the normal case. In one
variant, we swap a single star of mass $2m$ for each binary (``inert
binaries''). In the other, we swap 2 unbound single stars for each
binary (``all singles''). The suite of runs are summarized in
Table 2.

Three pair-wise comparisons may be made. To assess the full effect of
the presence or absence of binaries, we compare the normal run to the
one with all singles. The difference depends on at least two physical
effects present in the normal runs but absent in the runs with all
singles: mass segregation of the binaries and collisional encounters
between the binaries and the single stars. To gauge the effect of mass
segregation alone we compare the runs with inert binaries to those
with all singles. To judge the role of the binary internal degrees of
freedom alone we compare inert binary runs to normal runs.

Our results shows that the presence of binaries has only a small
influence on the overall dynamics.  Table 3 compares some key
quantities at $t=1.3$ for runs.  The relative differences in the total
external energy of the bound particles, the half-mass radius of the
bound particles and the mass of the bound particles are tabulated.
Averages of these quantities are given when multiple realizations of
the initial conditions were calculated. The first three
sections of the table show that only
small fractional differences have developed.  Repeated realizations
suggest that the differences are due to the random sampling
of the initial conditions; in any case, no systematic differences
between runs are observed.

The fourth section of
Table 3 provides quantitative information on the mass segregation
of actual and inert binaries. The final half mass radii of singles,
inert binaries and real binaries are compared and some systematic
trends emerge.  In all runs with the
inert binaries the heavy particles segregate compared to the singles.
However, in most runs with real binaries the binaries
are {\it less} concentrated than the singles, a consequence of the binary
destruction and the collisional recoil that occurs most rapidly at the
center. Clearly, the final spatial distribution of the binaries is as
strongly influenced by the collisional encounters as by the process of
mass segregation.

Table 4 examines changes in binary binding energy in the normal runs.
(Again average quantities are used where multiple realizations were
calculated.)  Although the {\it fractional} change in binary binding
energy is sizeable, the magnitude of that change is small compared to the
energy
carried off by escaping particles.  Taken together, the information in
Tables 3 and 4 shows that the mass and energy losses from ejected
particles dominate the changes caused by the presence of the binaries.
The systematic influence of binaries is small compared to the
intrinsic variations in escaping particles.

The maximum degree of contraction is a key factor influencing the rate
of collisional interaction of the binaries. Table 5 gives the ratio of
the initial half-mass radius $R_{h,i}$ to the half-mass radius at the
time of maximum contraction $R_{h,mc}$ for the runs with different
$N$. The best empirical fits to the degree of contraction are
powerlaws in the total number of particles $N_*=N_s+N_b$.\footnote{ We
use $N$ to indicate the total number of particles, $N=N_s+2N_b$, and
$N_*$ to stand for the ``effective'' number of particles
$N_*=N_s+N_b$. When we compare the results of section 2 to the
$N$-body simulations we use the quantity
$N_*$ rather than $N$ in the analytic formulae.} Figure 3
displays $\log R_{h,i}/R_{h,mc}$ as a function of $\log N_*$ (mean
values are used when multiple realizations were calculated).  The best
fit is
\be
 {R_{h,i} \over R_{h,mc}} = 1.13 N_*^{0.36},
\ee
in good agreement with the analytical and numerical estimate of
Aarseth et al. (1988) showing that $R_{h,i}/R_{h,mc} \propto N^{1/3}$.

{}From Tables 3, 4 and 5 we conclude (i) mass and energy loss is
dominated by ejected particles, (ii) the final structure of the bound
system is not grossly altered by the presence of binaries, (iii) the
maximum compression during violent relaxation is not greatly altered
by the presence of binaries. The particular choice of the form of
$f(\epsilon)$ does not alter these conclusions as long as single-star
binary interactions are more important than binary-binary
interactions. For our goal of exploring the binary truncation process
during violent relaxation, we are free to choose the form of
$f(\epsilon)$ that best elucidates the truncation mechanism, as we do
in the next section. We also conclude (iv) the spatial distribution of
the binaries is sensitive to the physical effects of mass segregation
and collisional interactions. Consequently, while (i)-(iii) are
independent of the particular choice of $f(\epsilon)$ that is not true
for the final spatial distribution of binaries.  For example, if $f(\epsilon)$
is dominated by high binding energy binaries, the
system's evolution tends to the case of the inert binaries in which
unambiguous mass segregation was evident.

The conclusions should not be significantly altered if $\xi$ is
changed. The initial conditions are very cold (small $\xi$), in the
sense that discreteness rather than initial thermal
dispersion governs the maximum compression during relaxation. The
cold initial conditions make the epoch of relaxation particularly
violent and maximize the collisional effects.  If warmer initial
conditions were used, particle ejection would decrease, total mass and
energy losses would diminish and the binary internal energy changes
would become relatively more important. As far as the collisional
processes are concerned, warmer initial conditions give an early epoch
that more closely resembles the quasi-stationary stage of evolution
discussed in \S 3.

\subsection{Ionization and hardening of the binary population}

The changes to the number of binaries $N_b$ and the binding energy
distribution are the focus of this section. The initial conditions
contain a range of binary binding energies $[\epsilon_{min},
\epsilon_{max}]$. At later times, a binary is taken to be any pair of
particles with total binding energy exceeding $\epsilon_{min}$.

Figure 4 shows the time evolution of $N_b(t)/N_b(0)$ for runs with
varying total particle number $N=1000$ (a), $2000$ (b), $3000$ (c) and
$5000$ (d). (Mean values are plotted for cases with multiple
realizations.) Most of the destruction evidently occurs close to the
instant the minimum radius is reached $1.1 < t < 1.2$.  It also of
interest that some binary destruction begins {\it before} that point.
The runs with larger $N$ suffer less of this early destruction.

Figure 5  shows the distribution of binding energies at the beginning
(solid lines) and at the end of the simulation ($t=1.3$, dashed lines)
for the different values of $N$. It is quite evident that the softest
binaries in the distribution suffer the greatest destruction. At the
same time, it is also clear that some binary hardening occurs; in
particular, all the final states include binaries with $\epsilon >
\epsilon_{max}$.

Although all the binaries included in the above simulations have
initially circular orbits, none of our essential results are altered
by considering a distribution of eccentricities. In fact, if
ionization is mainly responsible for the observed destruction, as we
will show below, no alteration is expected because the ionization rate
depends very weakly on eccentricity (Hut \& Bahcall 1983).  As a test
of this point we have carried out two additional simulations with
$N=1000$ (same initial virial ratio and binary fraction as the normal
runs) but with all binaries having initially eccentric orbits
($e=0.5$). In Figure 6 we compare the final distribution of binding
energy of binaries from these runs with that resulting from
simulations with binaries having initially circular orbits. No
significant differences are evident.

Figures 7 shows the time evolution of the hardness $x$ calculated
using the numerically determined central velocity dispersion (the
maximum value anywhere within the system.)  The vertical extent of the
line segment indicates the range of the hardness implied by the {\it
initial} binary binding energy distribution. The figure shows that
during the period before the phase of maximum contraction, the
binaries are hard so, in a thermal distribution, one would expect them
to harden, not disrupt. As a result, the early binary destruction
(when hardening is expected) is noteworthy. Moreover, after the
transition all the binaries are soft and, in a thermal distribution,
would tend to become softer. As a result, the binary hardening
observed at the end of the simulation for the largest initial binding
energies is noteworthy.

The data on the hardening prompts a question: how does one explain the
degree of hardening that is seen?  Although the positions of single
stars and binaries are randomly picked, the cold initial conditions
guarantee that correlations between particles are ``frozen in''; the
system must evolve for $\sim t_{ff}$ before the rate of hardening can
be calculated with the thermally averaged rate coefficients. In fact,
the binary hardening inferred using thermally averaged rate
coefficients (Heggie \& Hut 1993) and the numerically determined
density and velocity dispersion history is completely insufficient to
explain the observed binary hardening {\it before} the period of
maximum contraction or to explain the {\it total} binary hardening
observed by the end of the simulation.

It appears that part of the explanation is that there exist bound
triples present in the cold initial conditions. At $t=0$ some binaries
are bound to their closest neighbor; the hardening of such bound
systems cannot be described using the thermally averaged rate
coefficients. To show that such systems are expected, consider
replacing each binary with a single star with mass $2m$ located at the
center of mass of the binary. Write the total energy of the system
composed of the binary and the nearest neighbor:
\be
E_{triple}={1\over 2} \mu v^2-{Gm_1m_2\over d},
\ee
where $m_1=M_i/N$ is the mass of the single particle, $m_2=2m_1$ is
the mass of the binary, $\mu=(2/3)m_1$ is the reduced mass of the
system, $d=(4\pi/3)^{1/3}R_i/N^{1/3}$ is the initial average
interparticle separation while $v^2\simeq 1.2\xi{GM_i/R_i}$ is the
initial mean velocity dispersion.  We find
\be
E_{triple}={GM_i^2\over R_i}\left({0.4 \xi \over N}-{1.24\over N^{5/3}}\right)
\label{eq:triple}
\ee
which for $\xi=0.01$ and $N \sim 10^3$, as in our simulations, is
negative.  The triple begins near apocenter; if half the triple's
period is less than $t_{ff}$, pericenter will be reached before the
point of maximal contraction. If the minimum distance between the
binary and the single is less than the semi-major axis of the binary,
a strong interaction and binary hardening is a likely outcome.  We
estimate $t_{triple}$, half of orbital period of the single around the
binary star, to be
\be
t_{triple}\approx 2.1t_{ff}.
\ee
It is plausible that some binaries have a close interaction before
$t_{ff}$ and this is probably responsible for the hardening that takes
place at early times. These encounters cannot cause
ionization because the total energy of the triples is negative.

As check of the above arguments we have followed the binaries in the
run with $N=3000$. The number of binaries initially present in this
run is $N_b=150$; $70$ are initially bound to the closest particle to
form a triple system. The cumulative distribution function of the
distance of the distance to this closest particle, $d_{cp}$, is shown
in Figure 8a at three different times before maximum collapse ($t=0$,
$0.5$ and $1.$). As expected, the distance contracts as the system
collapses. In Figure 8b all the values of $d_{cp}$ at $t=0.5$ and
$t=1.0$ have been rescaled by the factor $R_{h,i}/R_h(t)$. About 80\%
of the triples collapse more quickly than the system as a whole.  This
is roughly consistent with the cumulative distribution function
of $t_{triple}$ in Figure 8c
showing that $60\%$ of the triples initially present in the system
have $t_{triple}<t_{ff}$.
Thus, it appears that a sizable fraction of
triple configurations have the time to collapse, i.e. to reach
pericenter, before maximum contraction.

Figure 8d shows the cumulative distribution function of the ratio of
the semimajor axis of the binary, $a$, to the
pericenter of the weakly bound element of the triple, $p$.
Nine binaries satisfy $p<2a$ and $t_{triple}<t_{ff}$, conditions which
guarantee a strong interaction.  This
supports the hypothesis that {\it some} hardening occurs because of the
interaction of singles and binaries initially bound in triple systems
and that the hardening can occur before the bounce.

It is important to note that a triple is not necessarily destroyed at
$t \sim t_{ff}$.  Hierarchical triples with $p \lta 2.75-3.5a$
(Harrington 1972) are inherently unstable and,
left alone, will eventually interact. The loosely bound outer star is
expected to harden the binary when the interaction occurs. Also, if
the triple encounters a single star it may produce an interaction with
the binary member. It is possible, but unproven, that nearly all the
triples will suffer strong interactions.

When triples are abundant the initial configuration may be more
important for binary hardening than subsequent thermal single-star
binary encounters.  For a hard binary, the number of single-star
binary encounters with a minimum distance of separation $\lta a$ over
the course of the period of violent relaxation is $\sim N_b
a/R_{h,mc}$ where $N_b$ is the total number of binaries.  Since
$R_{h,mc} \sim R/N^{1/3}$ and since the binary binding energies in the
numerical runs have been scaled so as to keep $x$ in the final
virialized state fixed, $a \propto R/N$, it follows that the
number of binary-single encounters in the N-body calculations $\sim
N_b/N^{2/3}$. Let $N_3$ be the number of bound triples that ultimately
give a strong binary interaction. The ratio of triple encounters to
single-star binary encounters scales like $N_3 N^{2/3}/N_b$; if the
system is so cold that each binary is bound in a triple (eqn.
[\ref{eq:triple}]) and if each triple suffers a strong interaction,
then the ratio increases with $N$. In fact, the observed change in
binding energy increases with $N$ in the set of simulations.

In summary, the cold initial conditions include bound triples (and
possibly higher-order assemblies) that interact and produce hardening
in some binaries before the point of maximum contraction is reached.
Bound triples may dominate the hardening after a free-fall time.

\subsection{Cutoff binary binding energy}

We now investigate the quantitative determination of the cutoff binary
semi-major axis $a_{cut}$, or equivalently, the cutoff binding energy
$\epsilon_{cut}$. We first introduce an {\it empirical} determination
based on comparison of the initial and final cumulative distribution
function (CDF) of the binding energy. Figure 9 gives the CDFs (initial
and final) normalized by the initial total number of binaries. We
define the cutoff to be the crossover point between the two curves.
This definition of the cutoff is not the same that we have
used in \S 2 or \S 5 so we discuss its plausibility.

The cross over point identifies the value of the binding energy
$\epsilon_{cut}$ such that the total number of more tightly bound
binaries is the same as in the initial conditions ($\Delta N =
N_f(>\epsilon)-N_i(>\epsilon) = 0$).  {\it If} it were true that
ionization was the only process responsible for modifying the binary
binding energy distribution, this definition would not be very useful
because some destruction can occur at all energies so $\Delta N < 0$ for
all $\epsilon$. A more meaningful definition might involve a
significant relative change ($\Delta N/N_i(> \epsilon) = \hbox{const}$).

On the other hand, it is clear from the CDFs in Figure 9 that
ionization is {\it not} the only process modifying the properties of
the binary star population. Significant hardening occurs for large
$\epsilon$ (evident from the fact that the final CDF is much larger
than the initial CDF for $\epsilon > \epsilon_{cut}$). This hardening,
which we have discussed in the previous section, is caused in part by
the bound triples in the initial configuration. Such encounters do not
lead to ionization and the rate of hardening encounters does not
appear to be very sensitive to $\epsilon$. For soft binaries the rate
of thermal ionization scales like $a$ (and Figure 7 shows that $x < 1$
at the cluster center) so it is clear that ionization is most
effective at small binding energies. Hence, the numerically determined
crossover point occurs at a point where the rate of ionizing
encounters exceeds the rate of hardening encounters.

 The numerical results are well fit
by the following expression
\be
\epsilon_{cut} ={1.14\over   N_*^{1.19}} {GM_i^2 \over R_i}.
\ee
Since $a=GM_i^2/(2 N^2 \epsilon)$, it follows
\be
a_{cut} ={0.40\over  N_*^{0.81}} R_i.
\ee
This numerical result differs in two ways from the theoretical
estimate (based on the simple model in \S 2) which gave
\be
a_{cut} = { 0.13 \over N^{0.36}} R_i .
\ee
The ratio of the two results is $\sim 0.33 N^{0.45}$; for $N=1000$
the ratio is about $7$, for $N=5000$ about $15$. That the magnitudes
differ by such factors is understandable given the variety of approximations
of the simple model. The difference in scaling is, however, more of
a puzzle.

There appear to be two possible explanations and they are not mutually
exclusive. First, in the analytic estimate, the asymptotic form for
the ionization rate $\scrR (x)$ for a soft binary was used.  As the
system evolves, a binary goes from being hard to soft and the relevant
asymptotic description for $\scrR (x)$ changes. Details may be found
in Appendix A.  Second, from the discussion of the initial
correlations of the cold system, it is clear that
the frequency of occurrence and the binding energy of
the triples both depend upon $N$. Only if $N \gg 10^3$ is the nearest star
likely to be unbound to the binary. It is not easy to perform an a
priori estimate of the scaling when such effects are present. Given
these uncertainties (in addition to those connected with the
original definition of the cutoff)
we also provide a more detailed calculation of the change in the
full binary distribution function in the next section.

The 1-dimensional mean square velocity dispersion in the virialized cluster is
\be
v_f^2 \simeq {2\over 15} {GM_f \over R_{h,f}}
\ee
 derived by means of the virial theorem and taking  the total energy of
the system equal to
$\vert E\vert=GM_f^2/(5R_{h,f})$ (see, e.g., Binney \& Tremaine 1987)
where  $M_f$ is the final cluster mass and $R_{h,f}$
is the final half-mass radius.
The hardness of the cutoff binary is
\be
x_{cut} ={ \epsilon_{cut} \over m v_f^2} = 3.75 {R_{h,f} \over a_{cut} N} {M_i
\over M_f}.
\ee
and the scaling of $x_{cut}$ with $N$ depends not only on the variation
of $\epsilon_{cut}$ but also on the mass and energy ejection from the
system.  The values of $x_{cut}$ for the simulations done are $[0.445$,
$0.342$, $0.257$, $0.154]$ for $N=1000$, $2000$, $3000$,and $5000$,
respectively. Assume (i) that $M_f$ and $R_{h,f}$ are
well-determined. Table 5 shows that the mass and energy loss do not
vary greatly over the entire set of runs. And,
(ii) that
the analytic scaling for $a_{cut}$ governs for $N > 5000$. Then, the
best empirical fit for the cutoff hardness is
\be
x_{cut} ={ 43 \over N^{0.65} } .
\ee

Clearly, as $N$ increases, $x_{cut}$ decreases. For all cases
investigated, the binaries destroyed in the course of the collapse are
unimportant in terms of the energetics of the cluster and do not
modify the dynamics of the collapse nor the final state resulting from
it.

\section{Analytical estimate of the effects of ionization on the binary stars}

We begin with an estimate of how the binary distribution function
evolves if thermal ionization were the only operative process and if
no mass segregation occurred.

We tabulated the time evolution (from $t=0$ to $t=1.3$) of number
density, $n_s(t)$, and velocity dispersion, $v_s(t)$, of single
particles at a set of 11 Lagrangian mass radii, $r_M$,
from the N-body calculations.  We assumed the
binaries to be associated with fixed Lagrangian radii throughout the
simulation. Using the ionization rate (\ref{eq:ionrate}) we calculate
$P(\epsilon, r_M)$, the probability of destruction for a binary of
energy $\epsilon$ at $r_M$.  Figure 10 shows the function
$P(\epsilon, r_M)$ for the runs with $N=1000$.  The largest ionization
probabilities for each value of binding energy are, as expected, those
calculated at the innermost Lagrangian radii where $n_s$ and $v_s$
reach the highest values.

The final binding energy distribution may be approximated
\be
f(\epsilon, t=1.3)=f(\epsilon,t=0)[1-\delta(\epsilon)]
\ee
where
\be
\delta(\epsilon) = {\int P(\epsilon,r) n_b d^3r \over
                    \int n_b d^3r } .
\ee
Figure 11 shows the initial and the final distribution function for
runs with $N=1000$, $2000$, $3000$ and $5000$ (dots give the final
binding energy distribution function for the $N$-body simulation, the
dashed line gives the initial distribution and the solid line gives
the above estimate of the final distribution). The binary hardening at
the largest values of $\epsilon$ is evident from the points that lie
to the right of the initial range of the distribution.  Apart from
this effect, the agreement between the two calculations is
satisfactory. Thermal ionization is responsible for the large changes that
occur at low energies.

Next, we account for the effect of the triple configurations.
Figure 12 shows the initial and final CDF for the run with $N=3000$
from $N$-body data (solid lines) and the analytic estimate (dashed
line). For small values of binding energy ($\epsilon/ m_*
\sigma^2< 11 $) the analytic CDF is determined by thermal ionization, as above.
The agreement is good for binaries less bound than the crossover
point. For higher values of the binding energy we assume that each
binary undergoes a single hardening encounter with a mean change of
$\Delta \epsilon=0.4 \epsilon$ (Heggie 1975).  The resultant final CDF
agrees well with the numerical CDF.

Finally, we compare the calculation of $\epsilon_{cut}$ and the total
fraction of binaries ionized with the results of the N-body
simulations (Tables 6 and 7). To find $\epsilon_{cut}$, we've chosen
$\delta = 0.21$ so that the crossover point of the analytic and
numerical results agree in the run with $N=1000$. (Any value of
$\delta$ of order $1$ is {\it a priori} reasonable.) The cutoff energy
and the fraction of ionized binaries both follow in the analytic
model, once this choice has been made. Values are compared with the
values obtained directly from N-body simulations in \S 2. The
agreement is quite good and suggests that the cutoff energy and binary
ionization is controlled by thermally averaged ionization.

\section{Conclusions and Future Work}

We have investigated how a population of primordial binaries is
modified by an epoch of violent relaxation.  We have mostly focused on
very cold initial conditions because they give rise to the maximum
degree of contraction and, hence, the strongest collisional effects.
We have used a binary fraction $f_b = 0.05$, thought to be typical of
today's globular clusters.

With a combination of analytic and N-body calculations we find the
following:

\begin{enumerate}
\item There is no significant change in the gross
cluster properties at the end of
violent relaxation due to the interaction between internal (binary)
degrees of freedom and translational degrees of freedom.
\item There is a characteristic binding energy $\epsilon_{cut}$ such that
binaries with $\epsilon > \epsilon_{cut}$ are not significantly disrupted,
while for $\epsilon < \epsilon_{cut}$, ionization is the main destructive
process.
\item The truncation in the energy
distribution of the binary population occurs for hardness parameter
$x<1$, as calculated in the virialized system.
\item As the number of particles $N$ increases,
collisional effects become less important and the cutoff binding
energy decreases.
\item The rate of hardening of binaries with $\epsilon>\epsilon_{cut}$
observed in the $N$-body simulations is not described by the thermally
averaged rate coefficients. Cold initial conditions used for the
simulations mean some binaries are bound to the closest single
particle. An interaction between the single particle and the binary is
responsible for at least part of the observed hardening.
\end{enumerate}

The scaling of the cutoff energy and the cutoff semi-major axis with
the number of particles in the system may be related to the size of
the system at the moment of maximum contraction. Reasonable agreement
between $N$-body simulations and analytic estimates is found.

Future work may attempt to increase the fraction of binaries, to
incorporate a spectrum of masses, to include gas, and to consider
inhomogeneous initial conditions.

\section*{Acknowledgements}
We thank the referee Sverre Aarseth for his comments
that led us to extend and improve the manuscript.
D.F.C. acknowledges grants NAGW-2224;
E.V. acknowledges financial support from Scuola Normale Superiore and the
hospitality of the Department of Astronomy of Cornell University.
\newpage
\clearpage
\section*{References}
Aarseth S.J., Lin D.N.C., Papaloizou J.C.B., 1988, ApJ, 324, 288\\
Abt H.A., 1983, ARAA, 21, 343\\
Binney J., Tremaine, S. 1987, Galactic Dynamics (Princeton: Princeton Univ.
Press)\\
Duquennoy A., Mayor M., 1991, A\&A, 248, 485\\
Fall M.S., Rees M.J, 1985, ApJ, 298, 18\\
Gao B., Goodman J., Cohn H., Murphy B., 1991, ApJ, 370, 567 \\
Goodman J., Hut P., 1989, Nature, 339, 40\\
Harrington R.S., 1972, Celest. Mech., 6, 322\\
Heggie D.C., 1975, MNRAS, 173, 729\\
Heggie D.C., Aarseth S.J., 1992, MNRAS, 257, 513\\
Heggie D.C., Hut, P. 1993, ApJS, 85, 347\\
Hut P., 1983, 268, 342\\
Hut P., Bahcall J.N., 1983, ApJ, 268, 319\\
Hut P., McMillan S.L.W., Goodman J., Mateo M., Phinney E.S., Pryor C.,
Richer H.B., Verbunt F., Weinberg M., 1992, PASP, 104, 981\\
Jayaraman S., Chernoff D.F., 1992, unpublished report\\
MacLow M., Shull J.M., 1986, ApJ, 302, 585\\
McMillan S.L.W., Aarseth S., 1993, ApJ, 414, 200\\
McMillan S.L.W., Hut P., Makino J., 1990, ApJ, 362, 522\\
McMillan S.L.W., Hut P., Makino J., 1991, ApJ, 372, 111\\
Reipurth B., Zinnecker H., 1993, A\&A, 278, 81\\
Salmon, J.K. 1991, Ph.D. Thesis at Cal Tech\\
Shapiro P.R., Kang H.,  1987, ApJ, 318, 32\\
Sigurdsson S., Phinney E.S. 1994, preprint\\
Spitzer L., Hart M.H. 1971, ApJ, 164, 399\\
van Albada T.S., 1982, MNRAS, 201, 939\\
Vesperini E., Chernoff D.F., 1994, ApJ, 431, 231\\
\clearpage

\section*{Appendix A}

In deriving expression (\ref{eq:delta})
for $\delta(a)$ we have assumed that the cut-off
binding energy falls in the very
soft regime at the moment of maximum collapse of the system;
the scaling law (\ref{eq:acutoff}) for $a_{cut}$ that follows from
eq.(\ref{eq:delta}) is different from that obtained from $N$-body simulations.
An explanation may be due in part to the fact that if the cut-off
binding energy does not fall in the very soft regime at the moment of
maximum contraction, as we have assumed above, the approximate expression
for $\scrR(x)$ we have used is incorrect.
In the following we will show that depending on the hardness factor of the
cut-off binding energy at the moment of maximum contraction three different
scaling laws for $a_{cut}$ hold.
If we define
\be
\tilde a \equiv a/R_{h,mc}
\ee
we can write
\be
\delta(a) = C\tilde a^2 N \scrR \left(B/\left[\tilde a N\right]\right)
\ee
where $B$ and $C$ are numerical constants introduced in the text.

It is easy to show that for some range of values of $\tilde aN$,

\be
\delta(a) =
\cases { 5.56 {\tilde a},~~~~~~~~~~~~~~~~~~~\hbox{if}~~~~ \tilde a N
\rightarrow
\infty, \cr
0.47 \tilde a^2 N,~~~~~~~~~~~~~~\hbox{if}~~~~ \tilde a N \sim B,
\cr
1.9 \tilde a^2 N \hbox{e}^{-B/ \tilde a N},~~~~~~~\hbox{if}~~~~ \tilde a N
\rightarrow 0 \cr}
\ee

Thus it
is clear from the above considerations that, in relation to the spacing
between the solutions of the equation $\delta(a) = \hbox{const.}$ used to
define the cut-off binding energy (or semi-major axis) there exists
three different regimes:
\begin{enumerate}
\item $N  \rightarrow \infty$:
$\delta$ enters  the linear regime for very small values of $\tilde a$ and
does not depend on $N$  so that $\tilde a_{cut} = \hbox{const.}$
\item 'Intermediate' $N$: solutions
in the regime where $\delta(a) \sim \tilde a^2 N$ and thus
$\tilde a_{cut} \sim N^{-1/2}$.
\item Low $N$: a steeper decrease with $N$
than $\tilde a_{cut} \sim N^{-1/2}$ is expected.
\end{enumerate}

The $N$-body calculations, described in the text, with $N
\sim 10^3-10^4$ have a scaling for $\tilde a$ that
is close to the 'intermediate' N regime ($\tilde a
\propto N^{-0.45}$ in this range). If, in calculating the solution of
$\delta(a) = \hbox{const.}$, we choose the
value of the constant  ($\simeq 0.003$)   so to get the correct scaling
for the range of $N$ investigated, we obtain the following estimates for
$a_{cut}$
\be
a_{cut} =
\cases {0.19 R_{h,mc}N^{-0.65},~~~~~~~~~~~\hbox{for }~~~~ N<500 \cr
0.08 R_{h,mc}N^{-1/2}~~~~~~~~~~~~~~~~\hbox{for}~~~~ 500<N<2000,
\cr
5.39 \times 10^{-4}
R_{h,mc},~~~~~~~~~~~~~\hbox{for}~~~~  N \rightarrow \infty
 \cr}
\ee
For the range of $N$ spanned by numerical simulations ($950<N_*<4750$)
the curve of the solutions of the equation  $\delta(a) =
\hbox{const.}$ is well fitted by
\be
a_{cut}={0.05 \over  N_*^{0.44}}R_{h,mc} ~~~~~~~~~~~~~~~~\hbox{for}~~~~
950<N_*<4750 \label{lasteq}
\ee

In this way we can obtain the observed scaling but there are still two
difficulties: 1) in order to get the correct scaling in the range of
$N$ where it is observed in $N$-body simulations we have been forced
to choose a value of the constant that is very small and it is not
clear why such a small value should be connected with the cut-off
energy; 2) the numerical constant in eq.(\ref{lasteq}) is much smaller than the
value found from N-body simulations. The latter point is likely to be
due, at least in part, to the very approximate estimates for the
density and velocity dispersion used to calculate $\delta(a)$ and for
this reason a more accurate analytical estimate has been carried out
in \S 5 of the paper.
\clearpage

\section*{Figure Captions}
Figure 1 Mass density profiles for the run with $N=1000$ and  initial virial
ratio $T/\vert V \vert = 0.01$ (no binaries) at (a) $t=0.5$, (b) $t=1.13$,
(c)
$t=1.3$ and (d) $t=2$.\\
Figure 2 Energy budget for the test run with $N_s=500$ single particles and
$N_b=250$ binaries. Total external energy of bound particles (short dashed
line), total external energy of escaping particles (solid line), total
(internal)  binding energy of binaries bound to the system (dashed line)
total (internal binding) of escaping binaries (long dashed line)
(see text for definitions). Note that the internal binding energy of the
binaries is taken here with negative sign.\\
Figure 3 Ratio of the
initial half-mass radius to the half-mass radius at the time of maximum
contraction. The dashed line is the best-fit line and it is proportional to
$N_*^{0.36}$.\\
Figure 4a-d Time evolution of the number of binaries normalized to
the initial number of binaries. For the values of $N$ (1000, 2000) for which
two
simulations have been carried out the mean value and the semidispersion as
error
bar are plotted. a) $N=1000$; b) $N=2000$; c) $N=3000$; d) $N=5000$\\
Figure 5a-d Histograms of the binding energy at the beginning of simulation
(solid line) and at the end of the simulation (dashed line) (a) $N=1000$; (b)
$N=2000$; (c) $N=3000$; (d) $N=5000$.\\
Figure 6 Histograms of the binding energy at the end of the simulations with
$N=1000$ and binaries initially with circular orbits ($e=0$) (solid line) and
eccentric orbits ($e=0.5$) (dashed line). Dots have been put on the histogram
corresponding to the runs with  binaries with initial eccentricity $e=0.5$ to
distinguish the two curves where they overlap. \\
Figure 7a-d Time evolution of
the hardness factor $x$ calculated at the center of the system (see text for
the
definition) for the range of values of binding energy put in the initial
conditions of N-body simulations. The upper and the lower point of each bar are
the hardness factor of the upper and lower values of binding energy initially
put in the systems  (a) $N=1000$; (b) $N=2000$; (c) $N=3000$; (d) $N=5000$.\\
Figure 8a-d  (a) Cumulative distribution function of the distance, $d_{cp}$,
between the closest particle and the binary forming bound triple systems at
$t=0$ (solid line), at $t=0.5$ (short dashed line) and at $t=1$ (long dashed
line); (b) as 10a but with the distances $d_{cp}$ at $t=0.5$ and $t=1$ rescaled
by a factor $R_{h,i}/R_h(t)$; (c) Cumulative distribution function of
$t_{triple}$ (see text); (d) Cumulative distribution function of the ratio of
the semimajor axes of the binaries bound in triple systems to the pericenter
distances between the binaries and the singles in the triple systems at $t=0$.
All the data refers to the run with $N=3000$.\\ Figure 9a-d Initial (solid
lines) and final (dashed lines) cumulative distribution functions of the
binding
energy of binaries in the N-body simulations normalized to the initial number
of
binaries. (a) $N=1000$; (b)  $N=2000$; (c) $N=3000$; (d) $N=5000$.\\ Figure 10
Destruction probability $P(r_M,\epsilon)$ as function of binding energy and
Lagrangian radius (see text for the analytical derivation)\\
Figure 11a-d  Final  binding energy
distribution function from $N$-body simulations  compared with the final
binding
energy distribution function derived analytically (solid lines). Dashed lines
show the initial binding energy distribution function. (a) $N=1000$; (b)
$N=2000$;
(c) $N=3000$; (d) $N=5000$.\\
Figure 12 Initial and final cumulative
distribution function  from N-body simulation (solid lines) for the run with
$N=3000$ and the final cumulative distribution function  from analytical
estimates (dashed line) calculated as described in the text: 1) $\epsilon/
m_* \sigma^2<11$  ionization only; 2)  $\epsilon/m_* \sigma^2>11$ hardening:
initial energy shifted by an amount equal to the mean change in energy per
encounter $\sim 0.4\epsilon$.\\
\newpage
\clearpage
\begin{table}
\begin{center}
\begin{tabular}{|ccccccc|}
\multicolumn {7} {c} {  Table 1}   \\
\multicolumn {7} {c} {Summary of test results}   \\
\hline
$\xi$&
$E_i$&$E_f$&$\Delta
M/M_i$&$\rho_{c,f}/\rho_{c,i}$&$R_{h,f}/R_{h,i}$&$\vert \Delta E/E_i
\vert $\\    \hline
0.25&0.44&0.44&-&19.4&0.45&$5\times 10^{-4}$ \\
0.1&0.53&0.61&0.15&115.6&0.26&$3\times 10^{-4}$ \\
0.01&0.59&0.92&0.18&1227&0.17&$2 \times 10^{-4}$ \\
 \hline
\end{tabular}
\end{center}

\caption[Table1]{Summary of test results.
All runs have $M_i = 1$, $R_i = 1$ and $N=1000$. The final
time is $t=2$,
$\xi$ is the initial virial ratio $T/\vert V \vert$,
$E_i$ is the total initial energy,
$E_f$ is the total final energy of particles bound to the system,
$\Delta M/M_i$ is the fractional mass loss,
$\rho_{c,f}/\rho_{c,i}$ is the ratio of final to initial central densities,
$R_{h,f}/R_{h,i}$ is the ratio of final to initial half-mass radii,
and $\Delta E/E_i$ is the fractional change in energy due to computational
errors.}
\end{table}
\newpage
\clearpage
\begin{table}
\begin{center}
\begin{tabular}{|ccccccc|}
\multicolumn {7} {c} {  Table 2}   \\
\multicolumn {7} {c} {  Initial conditions for N-body }   \\
\hline \hline
& &\# runs&  & &&\\
$N$ & Normal &Inert Binaries&Singles  & $\epsilon/m \sigma^2$
&$\epsilon/m_* \sigma^2$&eccentricity \\   \hline
1000&2&2&2&5-50&5-50&0  \\
1000&2&0&0&5-50&5-50&0.5  \\
2000&2&2&2&5-52&2.5-26&0  \\
3000&1&1&1&5-60&1.66-20&0  \\
5000&1&1&1&5-60&1-12&0  \\
 \hline
\end{tabular}
\end{center}

\caption[Table2]{Initial conditions for the N-body simulations.
All runs have $M_i = 1$, $R_i = 1$ and an initial virial ratio $\xi = 0.01$.
$N$ is the number of stars. Normal runs include $0.9N$ singles and $0.05N$
binaries, inert binary runs include $0.9N$ singles and $0.05N$ singles (of
mass $2m$), and single runs include $N$ singles.  For runs with
binaries, the range of binding energies is given $\epsilon/m\sigma^2$
(and $\epsilon/m_* \sigma^2$), where $m$ is mass per particle and $m_*
= 10^{-3}$, and $\sigma=\sqrt{\xi {GM_i\over R_i}}$ is proportional to
the initial velocity dispersion.}

\end{table}
\newpage
\clearpage
\begin{table}
\begin{center}
\begin{tabular}{|cccc|}
\multicolumn {4} {c} {Table 3} \\
\multicolumn {4} {c} {Comparisons}   \\
\hline
$N$&
${\Delta E_{f,b-s}\over E_{f,b} }$&
${\Delta E_{f,b-ib}\over E_{f,b} }$&
${\Delta E_{f,ib-s}\over E_{f,ib} }$\\
\hline
1000&0.029&0.099&$-0.077$ \\
2000&0.012&$-0.087$&0.091  \\
3000&0.043&0.081&$-0.041$  \\
5000&0.014&0.139&$-0.145$ \\
\hline
$N$&
${\Delta R_{h,b-s}\over R_{h,b} }$&
${\Delta R_{h,b-ib}\over R_{h,b} }$&
${\Delta R_{h,ib-s}\over R_{h,ib} }$\\
\hline
1000&$-0.043$&$-0.064$&0.020 \\
2000&0.093&0.225&$-0.170$ \\
3000&0.112&$-0.041$&$0.147$  \\
5000&$-0.003$&$-0.153$&0.132 \\
\hline
$N$&
${\Delta M_{f,b-s}\over M_{f,b} }$&
${\Delta M_{f,b-i}\over M_{f,b} }$&
${\Delta M_{f,ib-s}\over M_{f,ib} }$\\
\hline
1000&$-0.004$&$-0.019$&0.015 \\
2000&0.005&0.036&$-0.032$ \\
3000&0.035&0.001&0.034 \\
5000&$-0.014$&$-0.012$&$-0.002$\\
\hline
$N$&
${(R_{h,f}(s)-R_{h,f}(ib))\over R_{h,f}(s) }$&
${(R_{h,f}(s)-R_{h,f}(b))\over R_{h,f}(s) }$&
\\
\hline
1000&$0.17$&$-0.15$& \\
2000&$0.23$&$-0.38$&  \\
3000&$0.16$&$-0.11$& \\
5000&$0.13$&$0.02$&  \\
\hline
\end{tabular}
\end{center}

\caption[Table3]{
The first three sections report fractional differences in the final
energy ($\Delta E_f$), half-mass radius ($\Delta R_h$) and mass
($\Delta M_f$) for particles remaining bound to the cluster at
$t=1.3$.  The parameters of the runs are listed in Table 2.  Each
column applies to a pair of runs identical except for their treatment
of the binary component.  The pairs are identified by the subscripts
``b'', ``s'', ``ib'' which refer to runs with normal binaries, with
all singles and with inert binaries respectively.

The fourth section reports the fractional differences in the half mass
radii of separate components in the same run.
$R_{h,f}(s)-R_{h,f}(ib)$ is the difference between the final half-mass
radius of singles and inert binaries in the runs with inert binaries.
$R_{h,f}(s)-R_{h,f}(b)$ is the difference between the final half-mass
radius of singles and binaries in the runs with normal binaries.}

\end{table}

\begin{table}
\begin{center}
\begin{tabular}{|ccc|}
\multicolumn {3} {c} {Table 4} \\
\multicolumn {3} {c} {Changes in binding energy}   \\
\hline
$N$&${\Delta E_B\over E_{esc}}$&${\Delta E_B\over E_{B,i}}$\\
\hline
1000&$-0.006$&$-0.23$ \\
2000&$-0.007$&$-0.29$ \\
3000&$0.0$&$-0.02$ \\
5000&$0.005$&$0.39$ \\
\hline
\end{tabular}
\end{center}

\caption[Table4]{
A comparison of average relative energy changes for particles bound to
the cluster at $t=1.3$ for runs with normal binaries.  $\Delta E_B$ is the
difference between final and initial internal energy (negative binding
energy) of binaries bound to the cluster, $E_{esc}$ is the total
external energy carried away by escaping particles, and $E_{B,i}$ is
the total initial binary binding energy.}

\end{table}
\newpage
\clearpage
\begin{table}
\begin{center}
\begin{tabular}{|ccccc|}
\multicolumn {5} {c} {Table 5}\\
%\multicolumn {5} {c} {Maximal Compression}\\
\hline
$N$&type& $ R_{h,i}/R_{h,mc}$&$ \vert \Delta M \vert /M_i$&$\vert \Delta
E\vert /E_i$\\
\hline
1000 &Singles& 13.4&0.20&0.71 \\
     && 12.3&0.19&0.60 \\
     &Inert binaries& 11.0&0.20&0.53 \\
     && 11.6&0.17&0.56 \\
     &Normal& 13.7&0.22&0.79 \\
     && 13.7&0.17&0.63 \\
2000 &Singles& 13.1&0.21& 0.68 \\
     && 14.7&0.22&0.93 \\
     &Inert binaries&17.3&0.23& 0.99 \\
     & &15.3&0.25&0.95 \\
     &Normal& 14.8&0.21&0.82 \\
     && 14.7&0.21&0.83 \\
3000 &Singles& 16.7&0.25&1.07 \\
     &Inert binaries& 17.2&0.23&1.00 \\
     &Normal& 19.1&0.23&1.17 \\
5000 &Singles& 18.4&0.24 &1.20\\
     &Inert binaries&18.4&0.27 &1.15\\
     & Normal& 23.8&0.28&1.51 \\
\hline
 \end{tabular}
\end{center}
\caption[Table5]{
The degree of compression, mass loss and energy change in the N-body
simulations.  $R_{h,i}/R_{h,mc}$ is the maximum ratio of initial to
half-mass radii over the course of the calculation.  $\Delta M$ is the
difference between the final and the initial mass and $\Delta E$ is the
difference between the final and the initial external energy of the
particles bound to the system.}
\end{table}

\newpage
\clearpage
\begin{table}
\begin{center}
\begin{tabular}{|ccc|}
\multicolumn {3} {c} {  Table 6}   \\
\hline
$N$&
$\epsilon_{cut}(\hbox{N-body})/m_* \sigma^2
$&$\epsilon_{cut}(\hbox{analytic})/m_* \sigma^2$
\\    \hline  1000&34 &34 \\
2000&14 &11.5 \\
3000&9 &9.9 \\
5000&5 &5.5\\
\hline
\end{tabular}
\end{center}
\caption[Table6]{
A comparison of the cutoff energy observed in the N-body simulations
and the analytic estimate.}
\end{table}
\newpage
\clearpage
\begin{table}
\begin{center}
\begin{tabular}{|ccc|}
\multicolumn {3} {c} {Table 7}   \\
\hline
$N$& $D$(N-body)&$D$(analytic) \\   \hline
1000&$0.48\pm 0.05$&0.51 \\
2000&$0.48\pm0.05$&0.37 \\
3000&0.42&0.39 \\
5000&0.41&0.40 \\
\hline
\end{tabular}
\end{center}
\caption[Table7]{
A comparison of the fraction of disrupted binaries (D) in the
N-body simulations and the analytic estimate.}
\end{table}
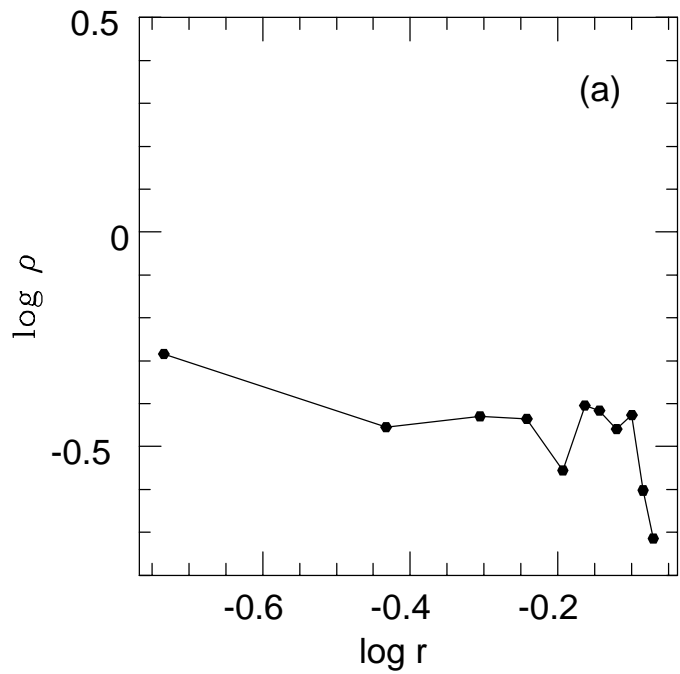
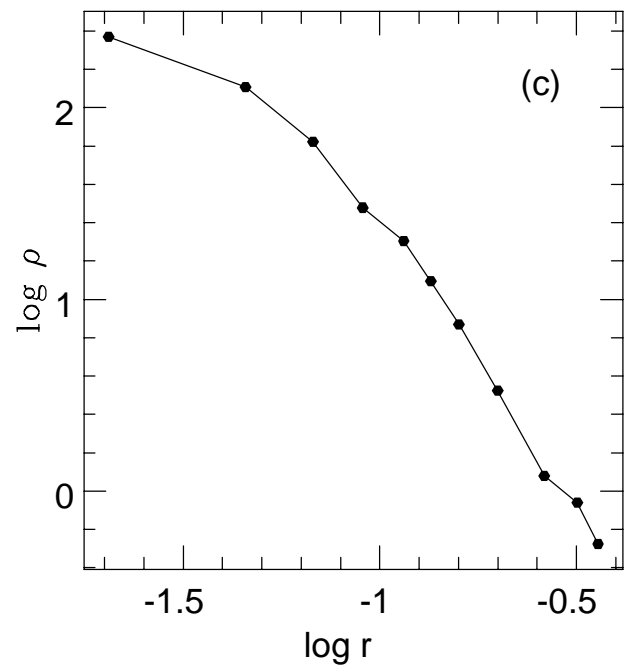
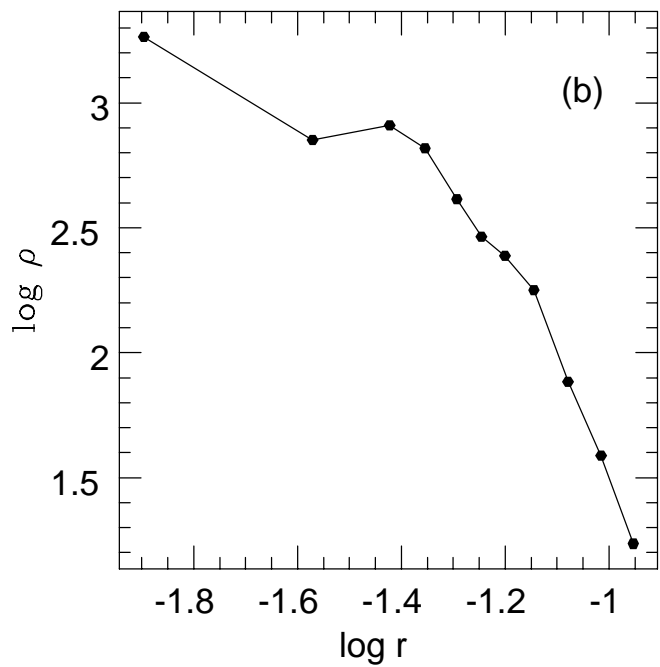
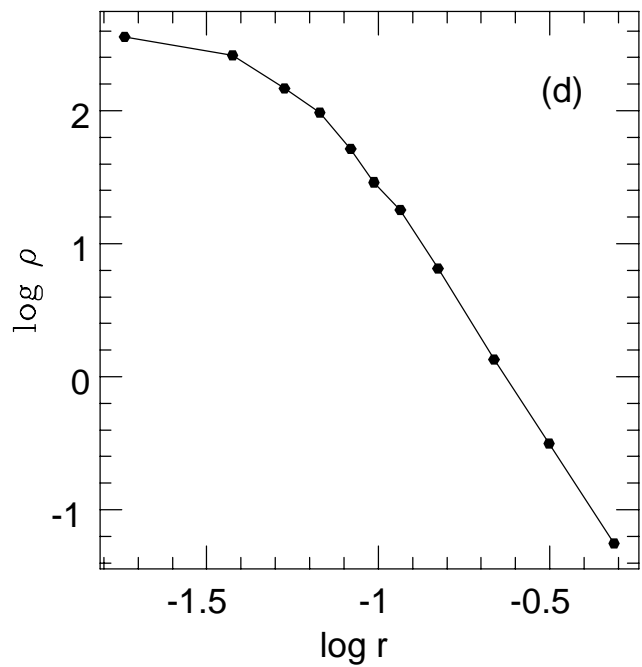
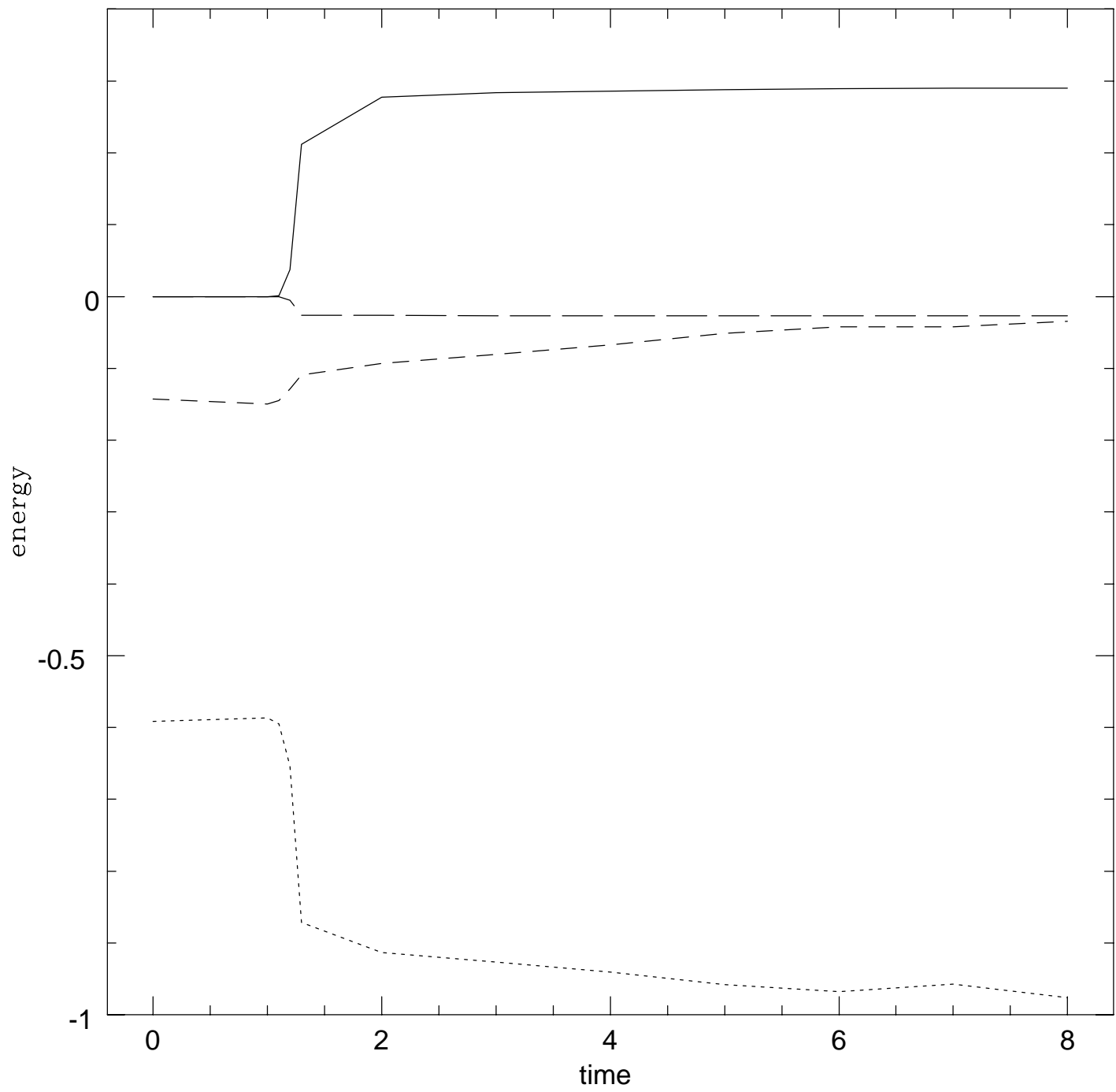
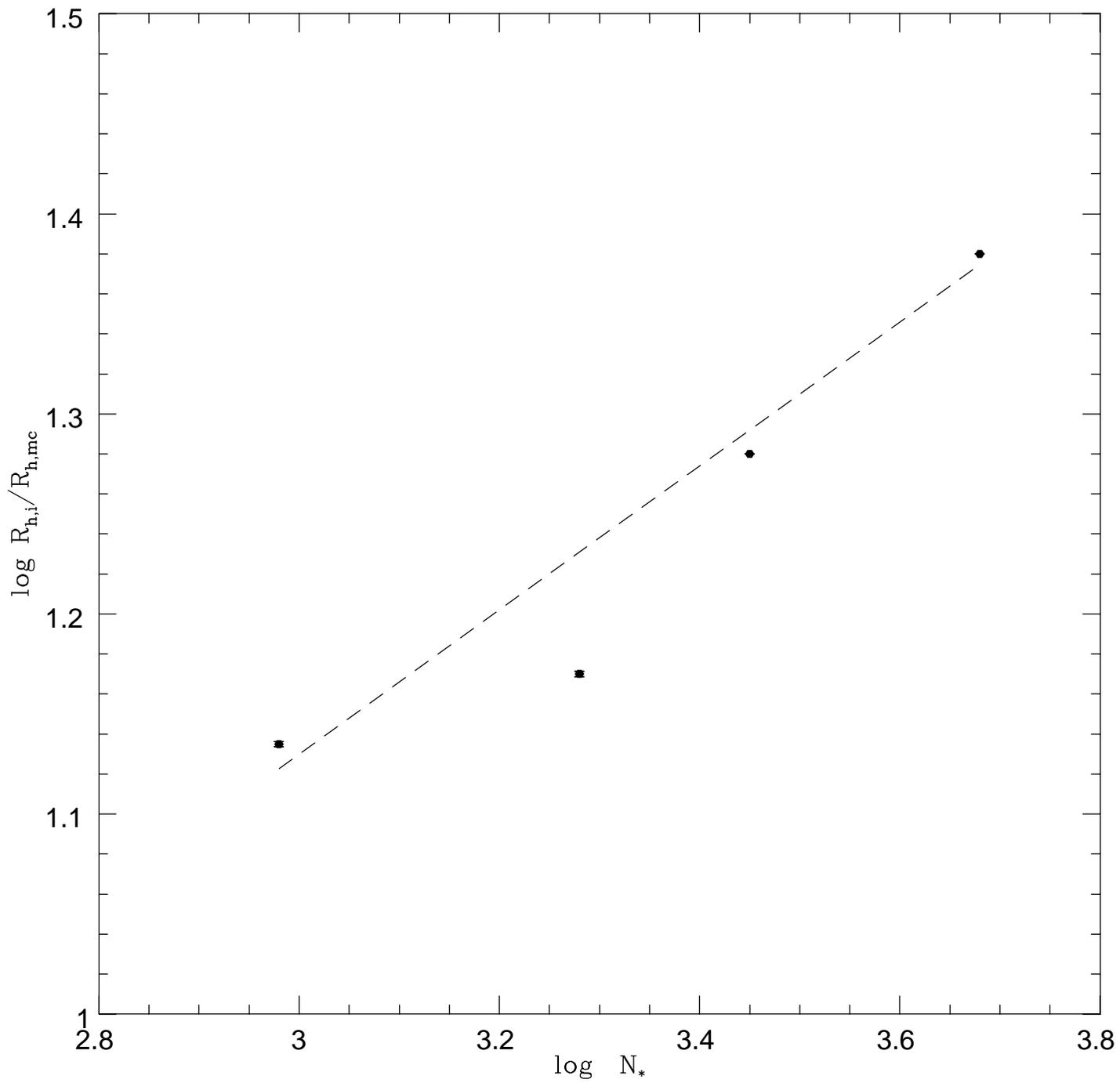
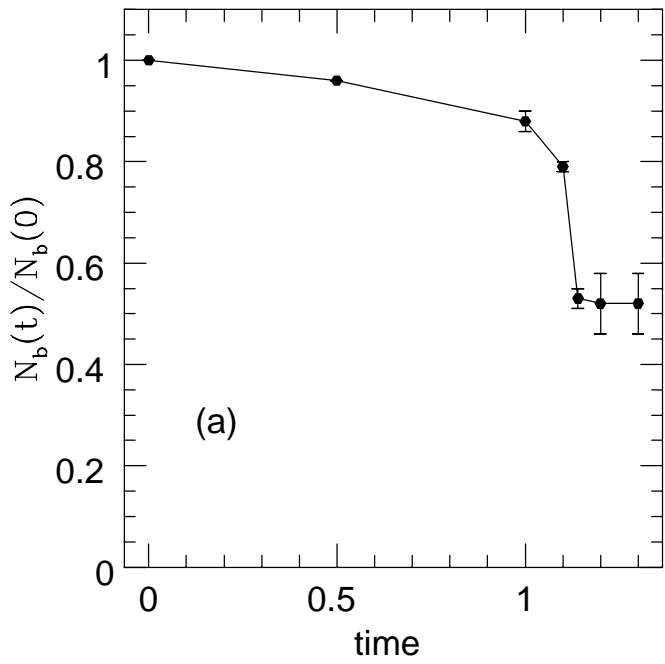
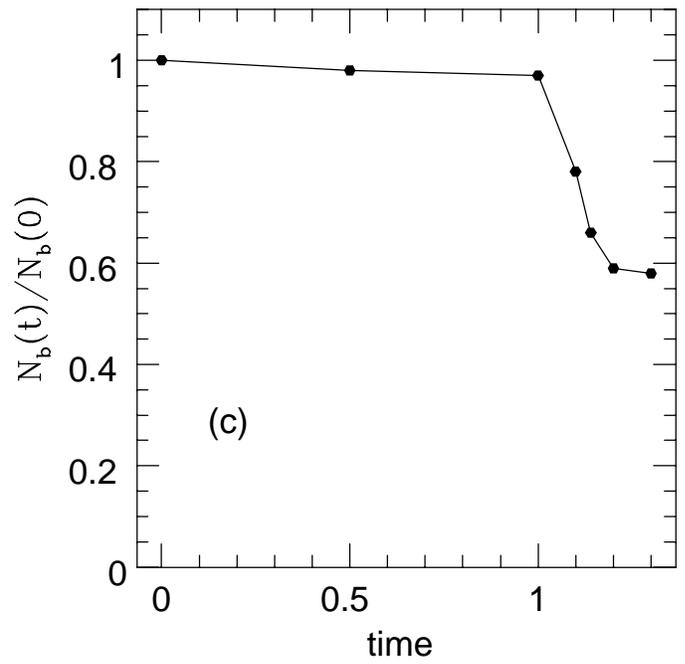
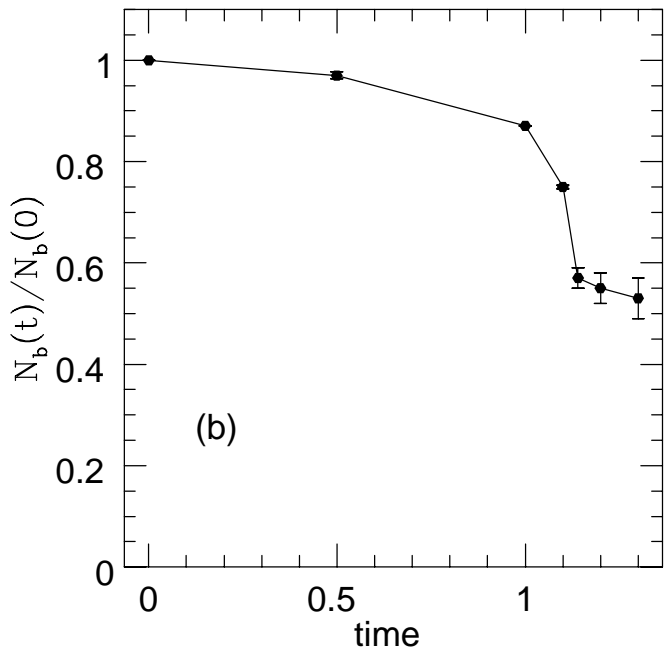
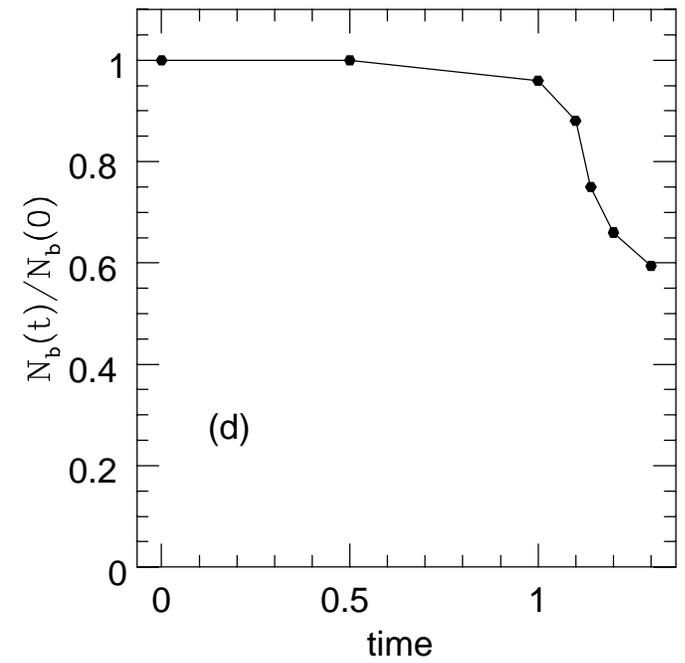
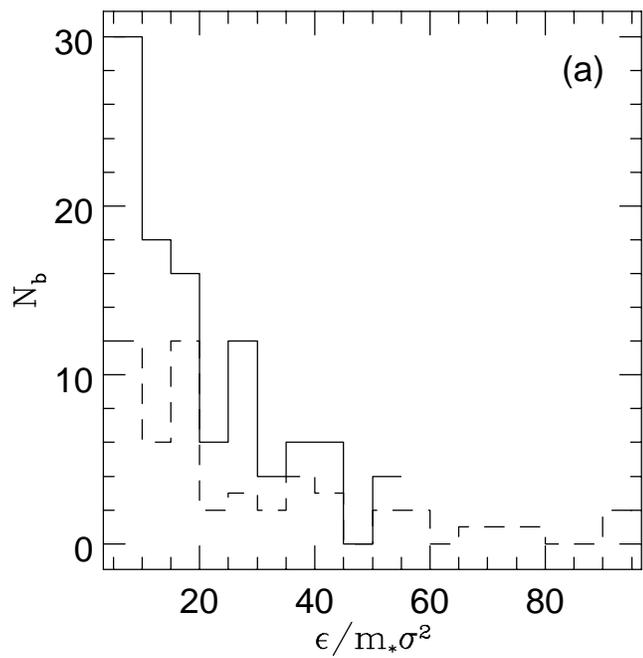
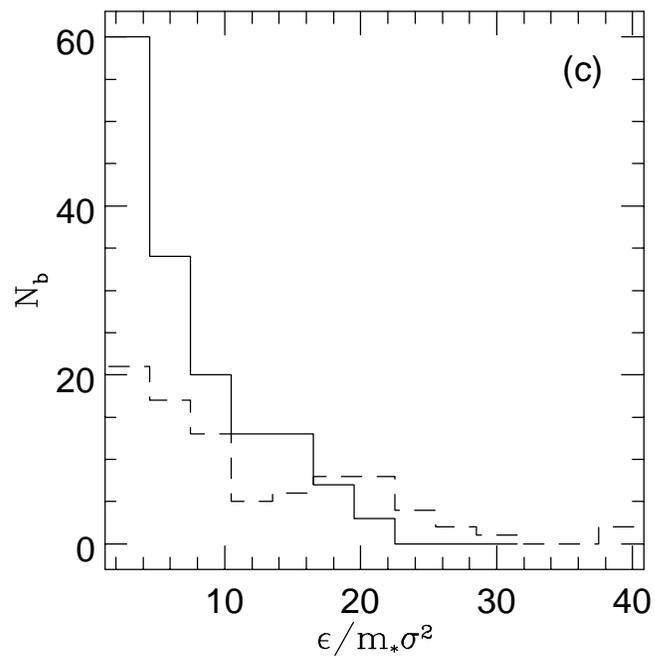
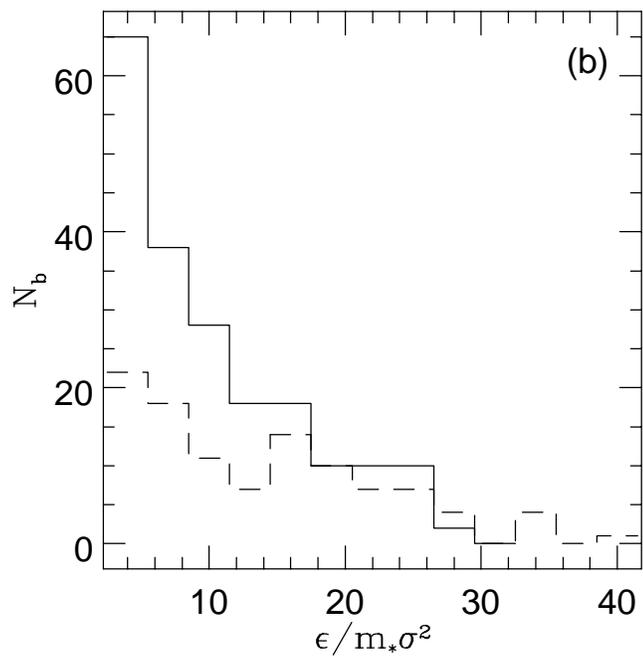
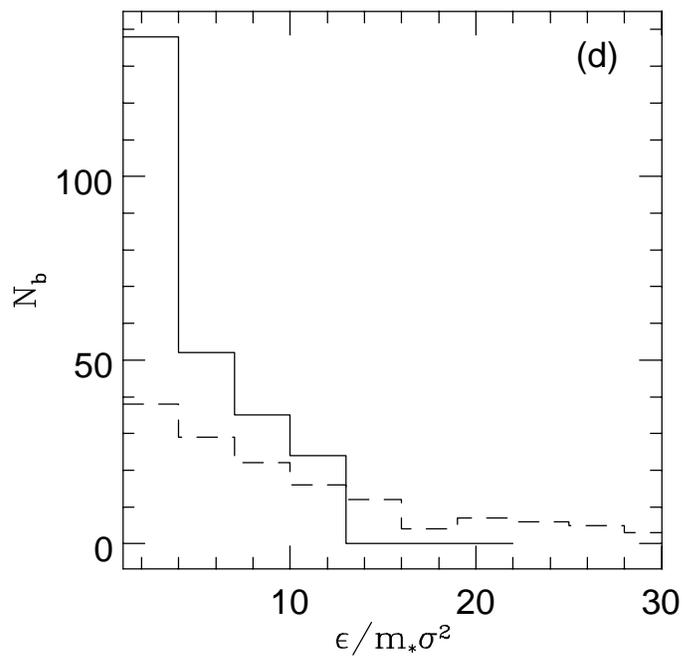
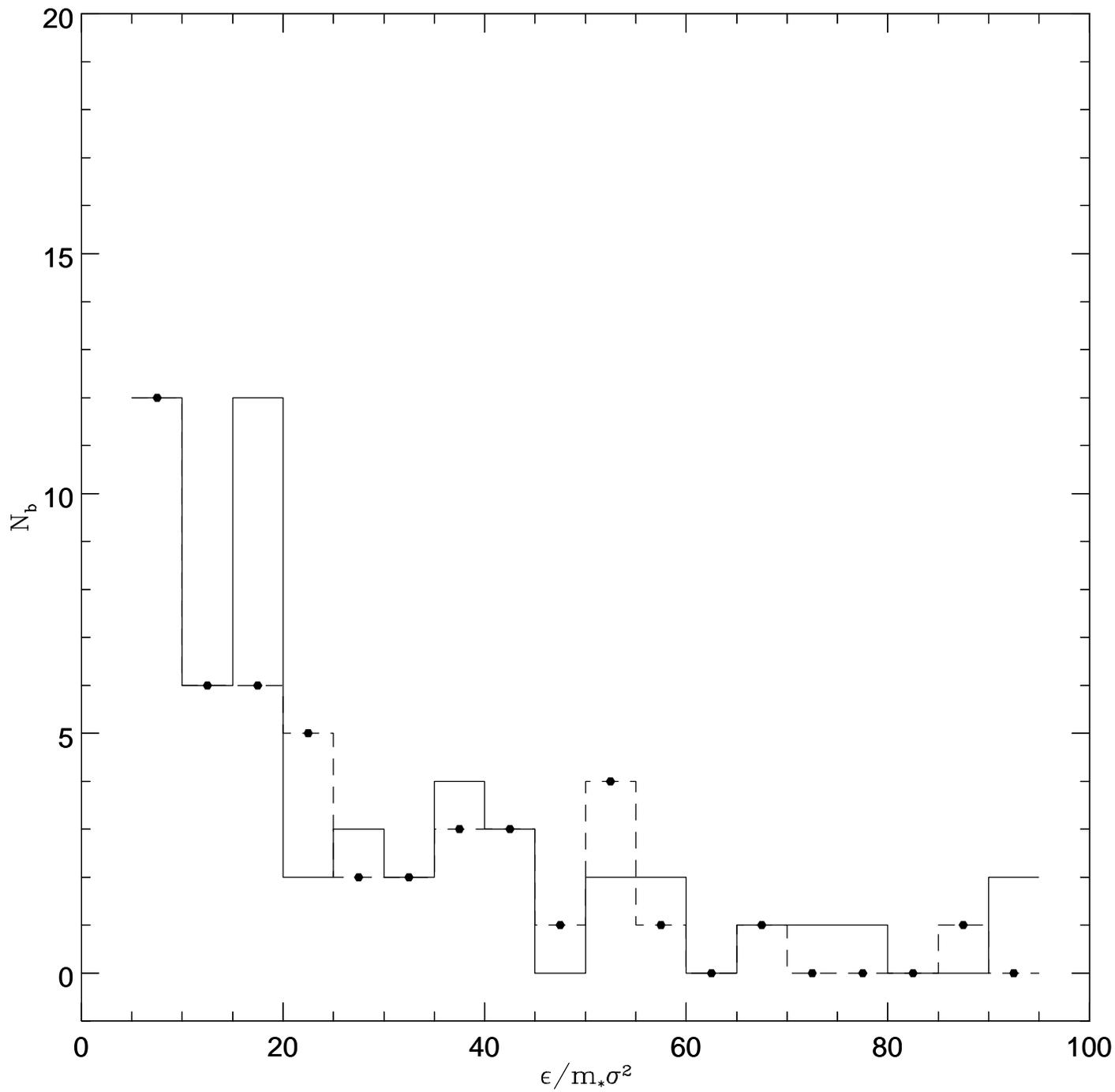
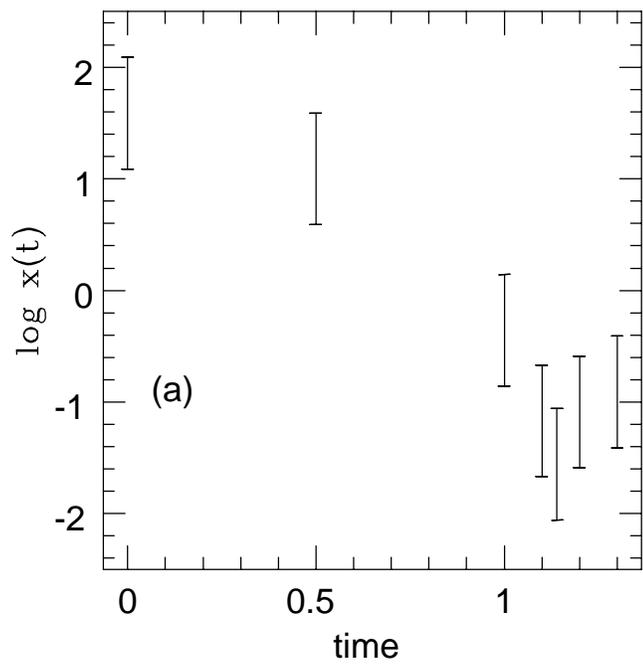
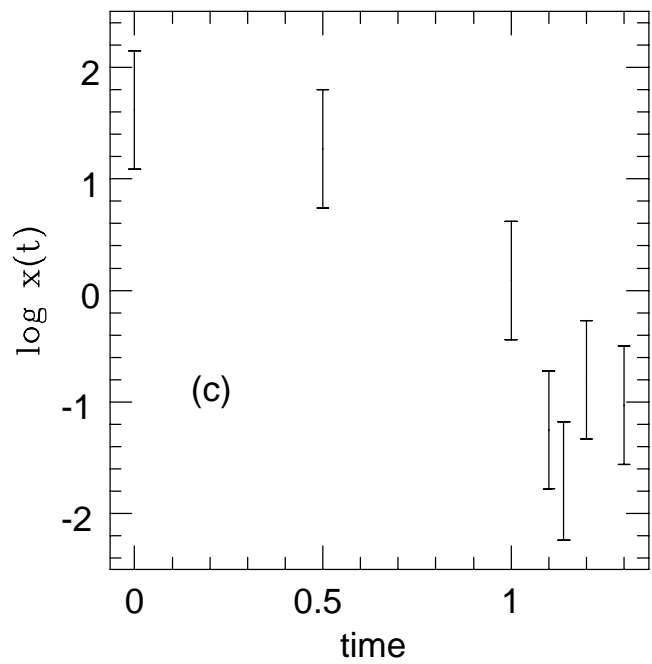
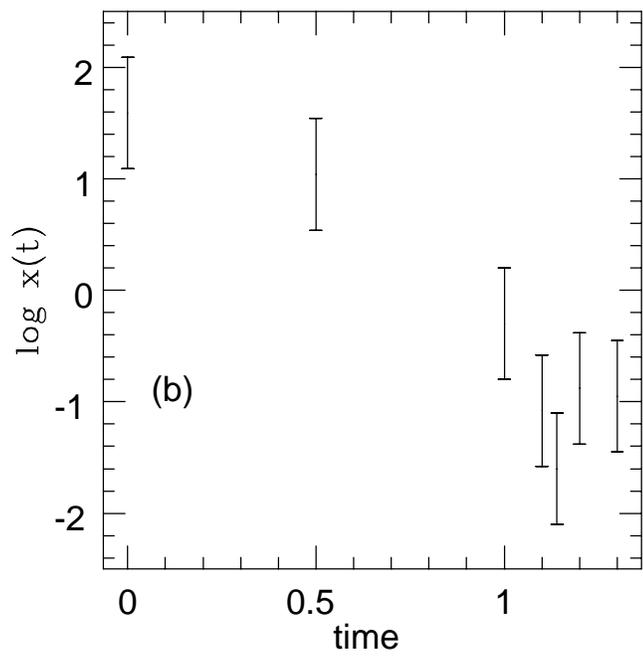
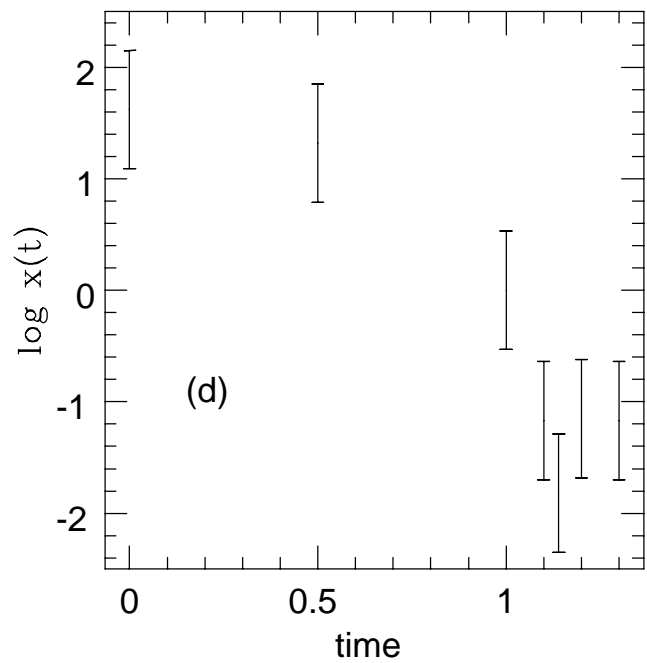
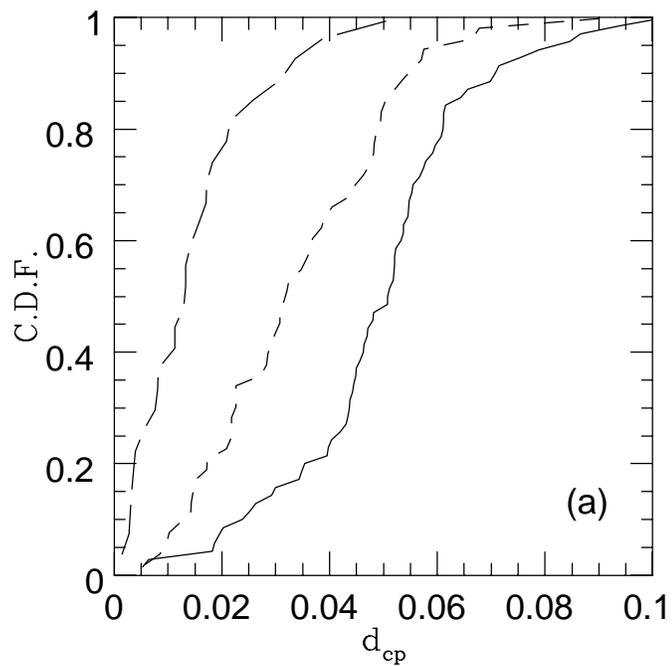
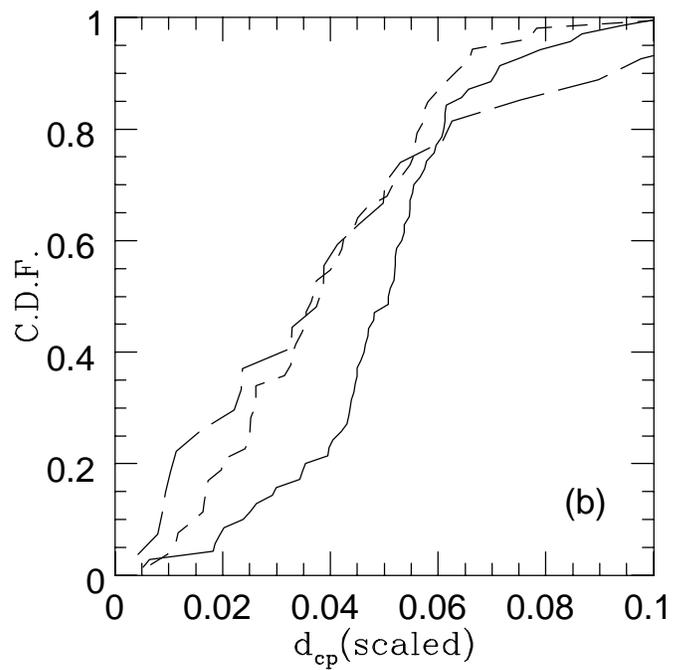
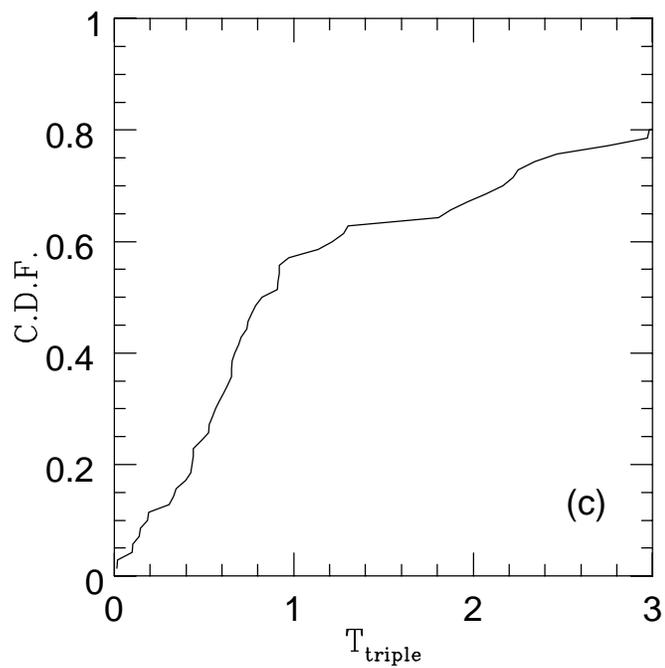
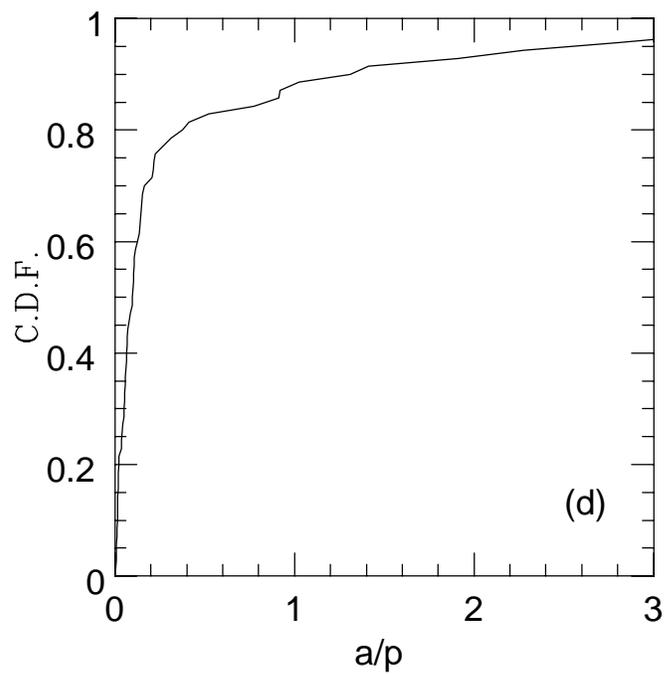
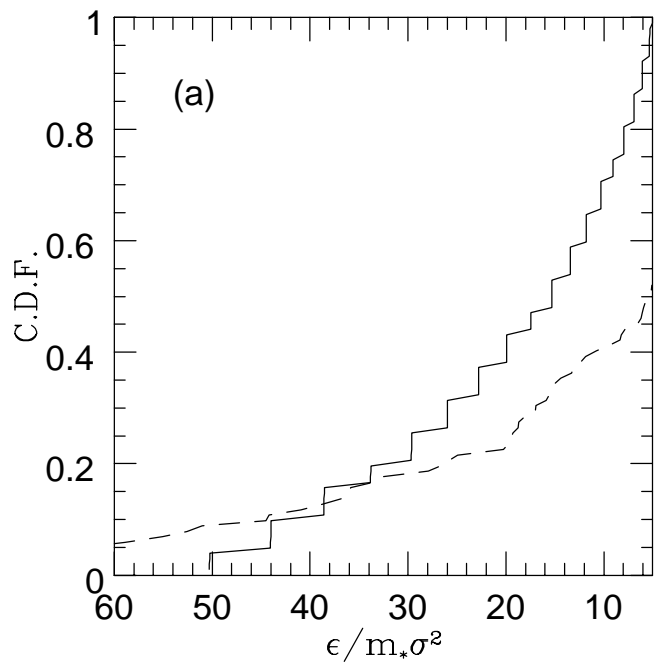
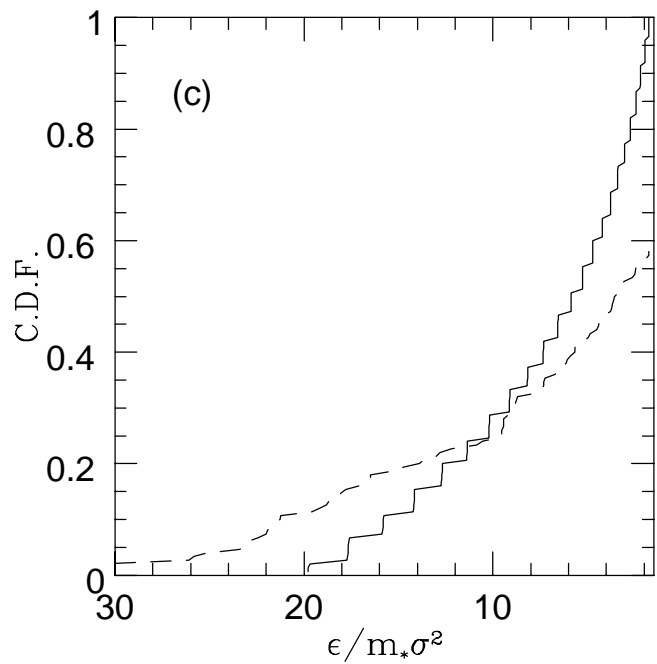
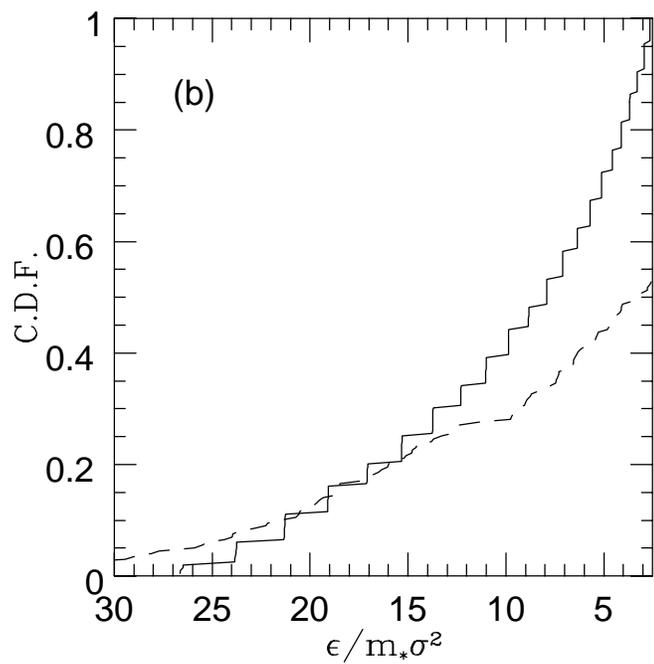
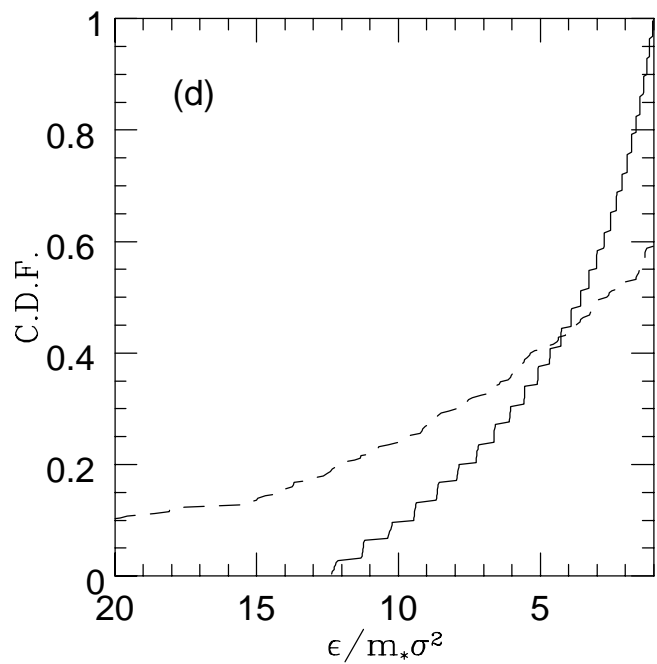
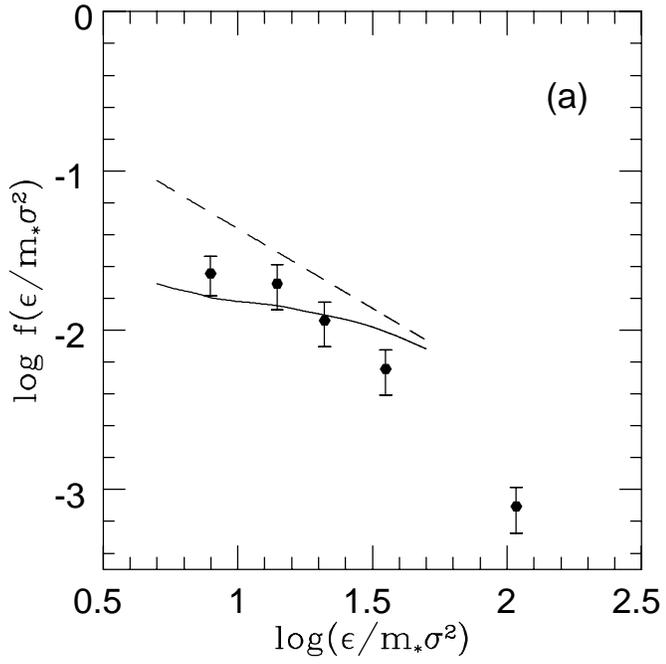
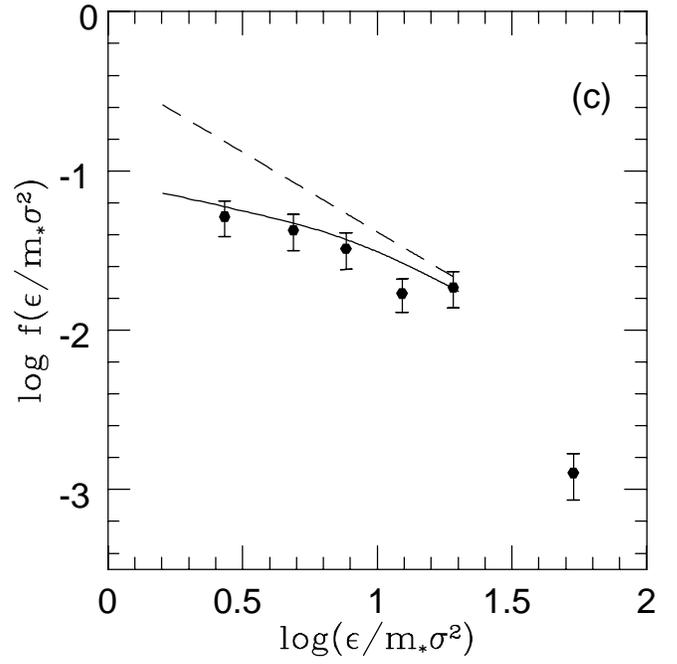
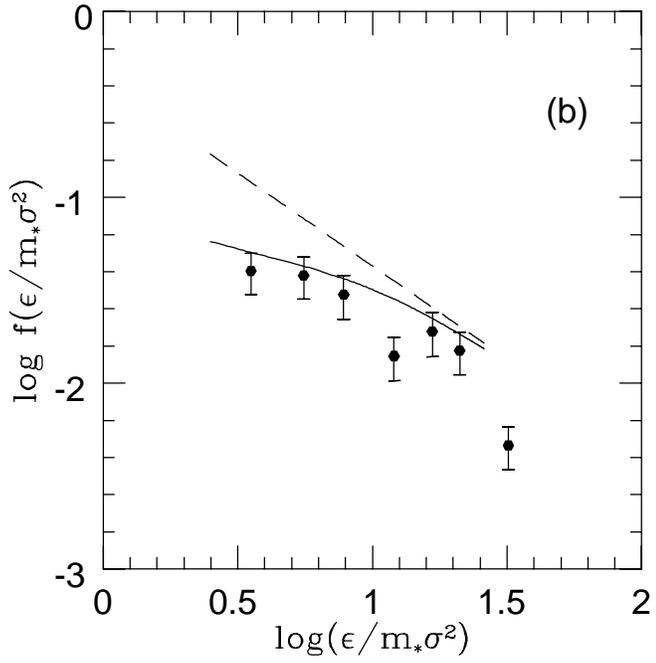
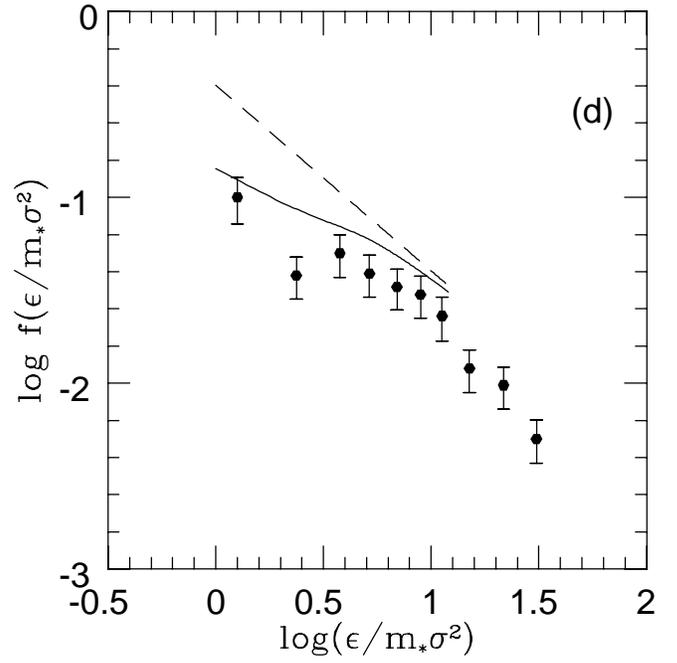
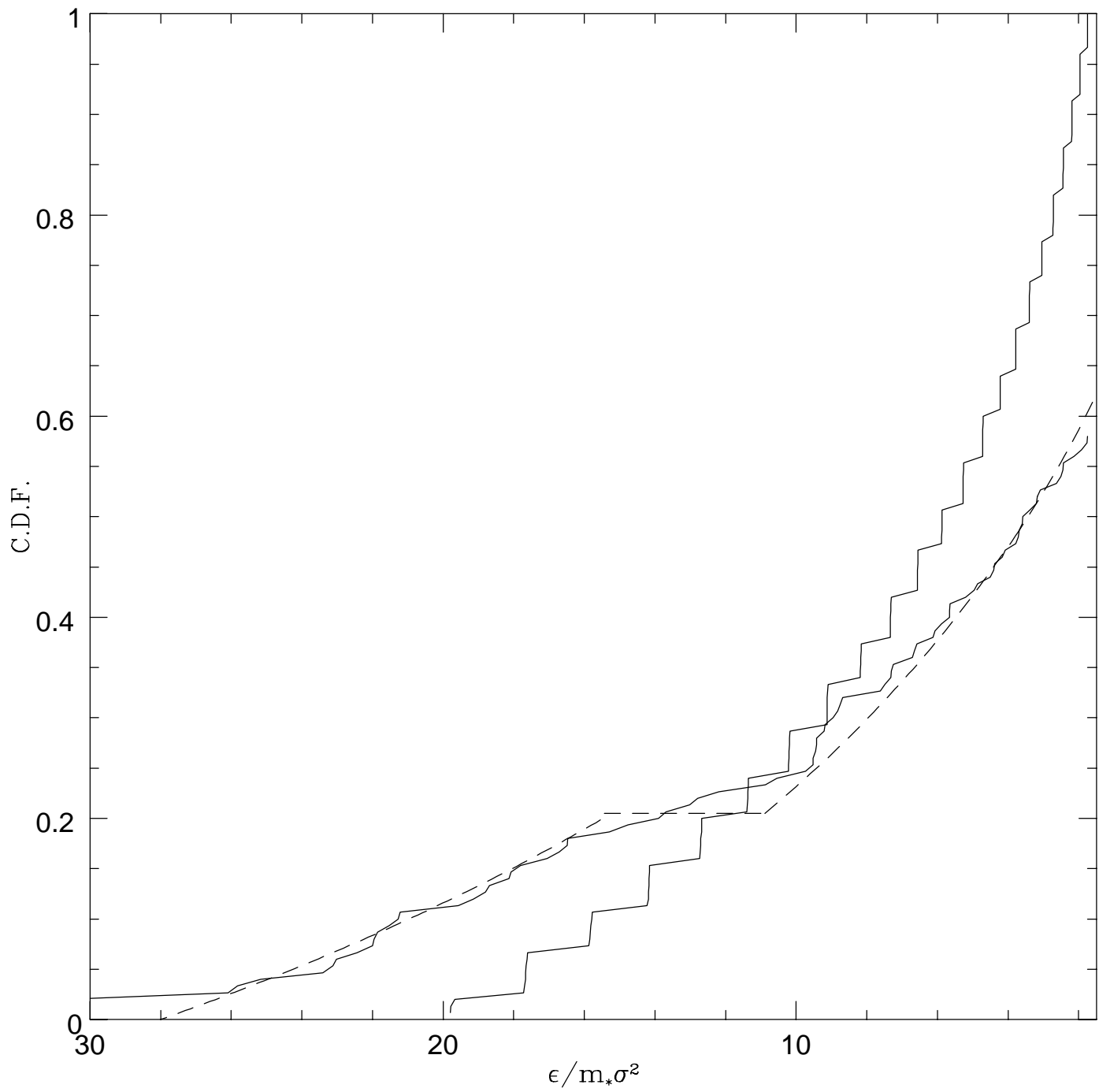
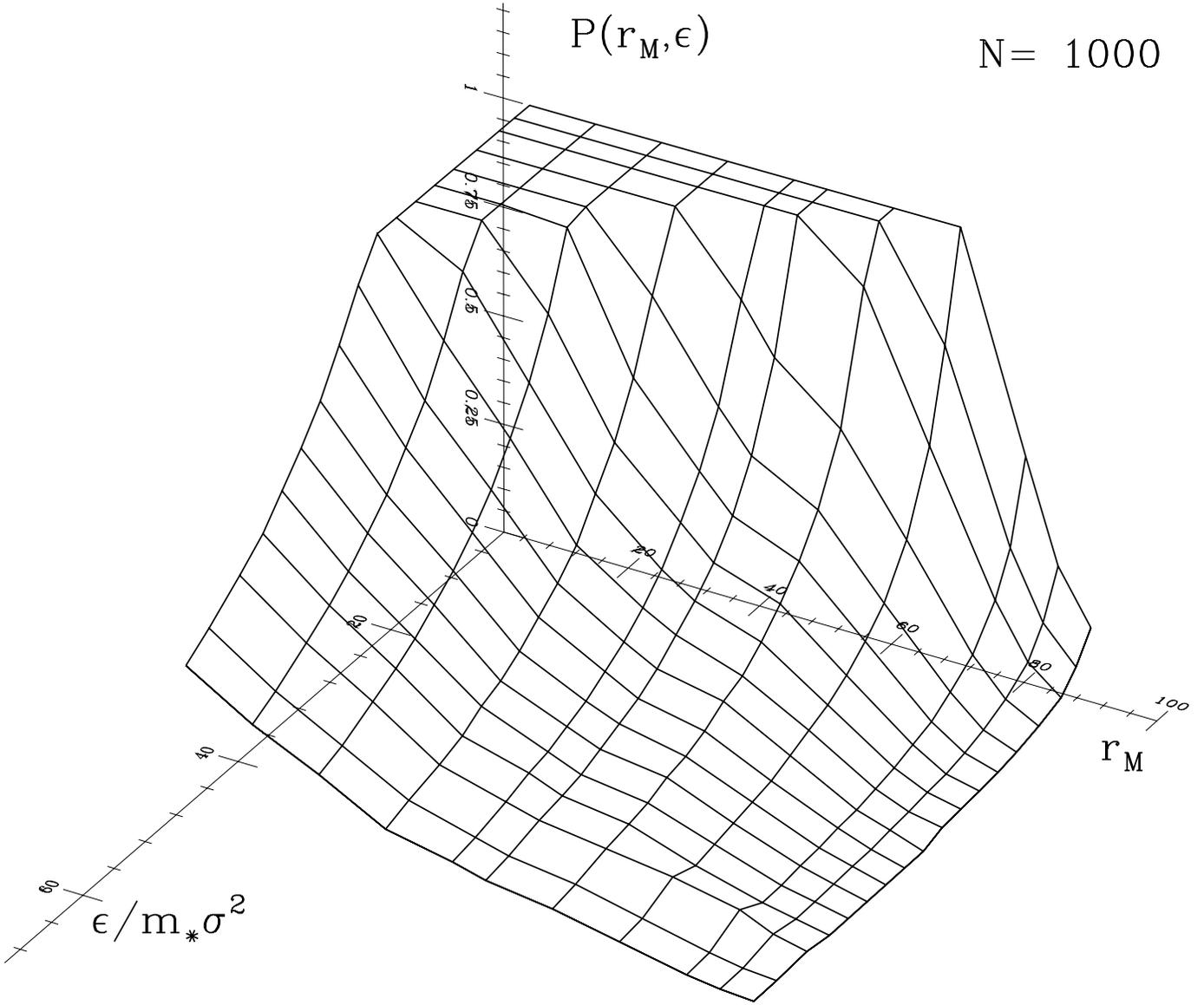
\end{document}